\documentclass[twocolumn]{aastex63}
\usepackage{natbib}
\def\code#1{\texttt{#1}}
\usepackage{multirow}
\usepackage{array}

\usepackage{amsmath} 
\usepackage{totcount}
\newtotcounter{citnum} 
\def\oldbibitem{} \let\oldbibitem=\bibitem
\def\bibitem{\stepcounter{citnum}\oldbibitem}
\usepackage{textcomp}
\usepackage{makecell}

\graphicspath{{./}{}}

\definecolor{notes}{HTML}{C70039}
\definecolor{softgreen}{HTML}{468465}
\definecolor{edits}{HTML}{BB8F00}

\newcommand{\oii}{\hbox{\sc [O\,ii]}}     
\newcommand{\hb}{\hbox{\sc H$\beta$}}       
\newcommand{\oiii}{\hbox{\sc [O\,iii]}}     
\newcommand{\ha}{\hbox{\sc H$\alpha$}}      
\newcommand{\nii}{\hbox{[N\,{\sc ii}]}}     
\newcommand{\sii}{\hbox{[S\,{\sc ii}]}}     

\newcommand{\lam}{$\lambda$}
\newcommand{\unit}[1]{\ensuremath{\mathrm{\,#1}}\xspace}

\newcommand{\ang}{$\mbox{\AA}$}

\newcommand{\kms}{km \,s$^{-1}$}

\newcommand{\e}{\unit{e^{-}}}

\mathchardef\mhyphen="2D

\newlength{\dhatheight}

\newcommand{\lenstool}{{\tt{Lenstool}}}

\newcommand{\LEGGOS}{LEGGOS} 
\newcommand{\SDSSeleventen}{SDSS\,J1110$+$6459}
\newcommand{\SDSSfiftintwentyseven}{SDSS\,J1527$+$0652}
\newcommand{\SDSSfourteentwentynine}{SDSS\,J1429$+$1202}
\newcommand{\SDSStwentyone}{SDSS\,J2111$-$0114}
\newcommand{\SDSStenfifty}{SDSS\,J1050$+$0017}
\newcommand{\SDSStwelvetwentysix}{SDSS\,J1226$+$2152}

\newcommand{\cosmiceye}{Cosmic Eye}
\newcommand{\sunburst}{Sunburst Arc}
\newcommand{\eleventen}{SGAS1110}

\newcommand{\twentyoneeleven}{SGAS2111}
\newcommand{\twelvetwentysix}{SGAS1226}

\newcommand{\JWST}{\textit{JWST}~}
\newcommand{\jwst}{\textit{JWST}~}
\newcommand{\HST}{\textit{HST}~}

\newcommand{\niib}{[\textrm{N}\textsc{ii}]\ensuremath{\lambda6584}}

\usepackage{fontspec}
\usepackage[english]{babel} 
\babelprovide{punjabi} 
\babelfont[punjabi]{rm}[
  Renderer=Harfbuzz,
  Script=Gurmukhi
]{FreeSerif.otf} 

\newcommand{\Punj}[1]{\selectlanguage{punjabi}{#1}\selectlanguage{english}}

\submitjournal{ApJ}

\shorttitle{\textit{LEGGOS}: JWST Spectra + Imaging of Clumps in Cosmic Noon Lensed Galaxies}

\begin{document}

\title{LEGGOS I: The \textit{JWST LEGGOS} Survey -- LEnsing and Galaxy Growth: Observing Substructures -- Unpacks the Nature of Clumpy Star Formation and Quenching in Gravitationally Lensed Galaxies beyond Cosmic Noon}


\author[0000-0002-3475-7648]{Gourav Khullar(\Punj{ਗੌਰਵ ਖੁੱਲਰ})}
\altaffiliation{Baum Postdoctoral Fellow for Innovative Astronomy}
\affiliation{Department of Astronomy \& the DiRAC Institute, University of Washington, Physics-Astronomy Building, Box 351580, Seattle, WA 98195-1700, USA}
\affiliation{eScience Institute, University of Washington, Physics-Astronomy Building, Box 351580, Seattle, WA 98195-1700, USA}
\affiliation{Department of Physics and Astronomy and PITT PACC, University of Pittsburgh, Pittsburgh, PA 15260, USA}

\author[0000-0001-5097-6755]{Michael Florian}
\affiliation{Steward Observatory, University of Arizona, 933 North Cherry Avenue, Tucson, AZ 85721, USA}
\affiliation{Eureka Scientific, 2452 Delmer Street Suite 100 Oakland, CA 94602-3017}

\author[0000-0003-1074-4807]{Matthew B. Bayliss}
\affiliation{Department of Physics, University of Cincinnati, Cincinnati, OH 45221, USA}

\author[0000-0001-6251-4988]{Taylor A. Hutchison}
\altaffiliation{NASA Postdoctoral Fellow}
\affiliation{Astrophysics Science Division, Code 660, NASA Goddard Space Flight Center, 8800 Greenbelt Rd., Greenbelt, MD 20771, USA}
\affiliation{Department of Astronomy, University of Maryland, Baltimore County, MD 21250, USA}
\affiliation{Center for Research and Exploration in Space Science and Technology, NASA/GSFC, Greenbelt, MD 20771 USA}

\author[0000-0003-1815-0114]{Brian Welch}
\affiliation{International Space Science Institute, Hallerstrasse 6, 3012 Bern, Switzerland}

\author[0000-0002-7559-0864]{Keren Sharon}
\affiliation{Department of Astronomy, University of Michigan, 1085 S. University Ave, Ann Arbor, MI 48109, USA}

\author[0000-0002-7627-6551]{Jane R. Rigby}
\affiliation{Astrophysics Science Division, Code 660, NASA Goddard Space Flight Center, 8800 Greenbelt Rd., Greenbelt, MD 20771, USA}

\author[0009-0000-5333-9970]{Dylan Berry}
\affiliation{Department of Astronomy \& the DiRAC Institute, University of Washington, Physics-Astronomy Building, Box 351580, Seattle, WA 98195-1700, USA}

\author[0000-0002-9204-3256]{T.\ Emil Rivera-Thorsen}
\affiliation{The Oskar Klein Centre, Department of Astronomy, Stockholm University, AlbaNova, 10691 Stockholm, Sweden}

\author[0000-0003-2200-5606]{H{\aa}kon Dahle}
\affiliation{Institute of Theoretical Astrophysics, University of Oslo, P.O. Box 1029, Blindern, NO-0315 Oslo, Norway}

\author[0000-0002-0108-4176]{Sedona H. Price}
\affiliation{Space Telescope Science Institute, 3700 San Martin Drive, Baltimore, MD 21218, USA}
\affiliation{Department of Physics and Astronomy and PITT PACC, University of Pittsburgh, Pittsburgh, PA 15260, USA}

\author[0000-0001-7151-009X]{Nikko J.\ Cleri}
\affiliation{Department of Astronomy and Astrophysics, The Pennsylvania State University, University Park, PA 16802, USA}
\affiliation{Institute for Computational \& Data Sciences, The Pennsylvania State University, University Park, PA 16802, USA}
\affiliation{Institute for Gravitation and the Cosmos, The Pennsylvania State University, University Park, PA 16802, USA}

\author[0000-0002-5293-3975]{Julissa Sarmiento}
\affiliation{Department of Physics and Astronomy, University of Pittsburgh, Pittsburgh, PA 15260, USA}

\author[0000-0003-3266-2001]{Guillaume Mahler}
\affiliation{STAR Institute, Quartier Agora - All\'ee du six Ao\^ut, 19c B-4000 Li\`ege, Belgium}

\author[0000-0003-1370-5010]{Michael D. Gladders}
\affiliation{Department of Astronomy and Astrophysics, University of Chicago, 5640 South Ellis Avenue, Chicago, IL 60637, USA}
\affiliation{Kavli Institute for Cosmological Physics, University of Chicago, 5640 South Ellis Avenue, Chicago, IL 60637, USA}

\author[0000-0001-5063-8254]{Rachel Bezanson}
\affiliation{Department of Physics and Astronomy and PITT PACC, University of Pittsburgh, Pittsburgh, PA 15260, USA}

\author[0009-0005-8103-5823]{Alex Ross}
\affiliation{Department of Astronomy \& the DiRAC Institute, University of Washington, Physics-Astronomy Building, Box 351580, Seattle, WA 98195-1700, USA}

\author[0009-0004-9243-3459]{Pedram Abedi}
\affiliation{Department of Astronomy, University of Michigan, 1085 S. University Ave, Ann Arbor, MI 48109, USA}

\author[0009-0000-7075-5554]{Rion Oh}
\affiliation{Department of Astronomy \& the DiRAC Institute, University of Washington, Physics-Astronomy Building, Box 351580, Seattle, WA 98195-1700, USA}
\affiliation{Department of Physics, KAIST, Daejeon 34141, Republic of Korea}

\author[0000-0002-0302-2577]{John Chisholm}
\affiliation{Department of Astronomy, University of Texas at Austin, 2515 Speedway, Austin, Texas 78712, USA}
\affiliation{Cosmic Frontier Center, The University of Texas at Austin, Austin, TX 78712, USA} 

\author[0000-0001-6505-0293]{Keunho Kim}
\affiliation{IPAC/Caltech, KS building, Office 310, IPAC at Caltech, Pasadena, CA 91125, USA}

\author[0000-0001-7160-3632]{Katherine E. Whitaker}
\affiliation{Department of Astronomy, University of Massachusetts, Amherst, MA 01003, USA}
\affiliation{Cosmic Dawn Center (DAWN), Copenhagen, Denmark}

\author[0009-0009-4672-7807]{Aleena Ebey}
\affiliation{Department of Physics, University of Cincinnati, Cincinnati, OH 45221, USA}

\author[0000-0002-4536-5463]{Cole Panzer}
\affiliation{Department of Physics, University of Cincinnati, Cincinnati, OH 45221, USA}

\author[0000-0002-0243-6575]{Jacqueline Antwi-Danso}
\altaffiliation{Dunlap Postdoctoral Fellow}
\affiliation{David A. Dunlap Dept. of Astronomy and Astrophysics, University of Toronto, 50 St. George Street, Toronto, M5S 3H4, Canada}

\author[0000-0002-8261-9098]{Catherine Cerny}
\affiliation{Department of Astronomy, University of Michigan, 1085 S. University Ave, Ann Arbor, MI 48109, USA}

\author[0000-0003-1343-197X]{Suhyeon C. Choe}
\affiliation{The Oskar Klein Centre, Department of Astronomy, Stockholm University, AlbaNova, 10691 Stockholm, Sweden}

\author[0000-0002-1728-8042]{Juliana S. M. Karp}
\affiliation{Department of Astronomy \& the DiRAC Institute, University of Washington, Physics-Astronomy Building, Box 351580, Seattle, WA 98195-1700, USA}

\author[0009-0000-3563-1695]{James W. Kulp}
\affiliation{Steward Observatory, University of Arizona, 933 North Cherry Avenue, Tucson, AZ 85721, USA}

\author[0000-0001-8367-6265]{Tim B. Miller}
\affiliation{Center for Interdisciplinary Exploration and Research in Astrophysics (CIERA), Evanston, IL 60201, USA}

\author[0000-0002-4606-4240]{Grace M.\ Olivier}
\affiliation{The Observatories of the Carnegie Institution for Science, 813 Santa Barbara Street, Pasadena, CA 91101, USA}

\author[0000-0002-2862-307X]{M. Riley Owens}
\affiliation{Department of Astronomy, University of California, Berkeley, Berkeley, CA 94720, USA}

\author[0000-0003-4075-7393]{David J. Setton}\thanks{Brinson Prize Fellow}
\affiliation{Department of Astrophysical Sciences, 4 Ivy Lane, Princeton University, Princeton, NJ 08544, USA}

\author[0000-0003-4702-7561]{Irene Shivaei}
\affiliation{Centro de Astrobiologia, Calle Ajalvir, 28864 Torrejon de Ardoz, Madrid, Spain}

\author[0000-0003-3302-0369]{Erik Solhaug}
\affiliation{Department of Astronomy and Astrophysics, University of Chicago, 5640 South Ellis Avenue, Chicago, IL 60637, USA}
\affiliation{Kavli Institute for Cosmological Physics, University of Chicago, 5640 South Ellis Avenue, Chicago, IL 60637, USA}

\author[0009-0003-7031-2907]{Amritaansh Srivastava}
\affiliation{Steward Observatory, University of Arizona, 933 North Cherry Avenue, Tucson, AZ 85721, USA}

\author{Sierra Bet}
\affiliation{Department of Astronomy \& the DiRAC Institute, University of Washington, Physics-Astronomy Building, Box 351580, Seattle, WA 98195-1700, USA}

\correspondingauthor{Gourav Khullar/Michael Florian}
\email{gkhullar@uw.edu/florianm@arizona.edu}

\begin{abstract}

We present first results from the \textbf{\JWST LEGGOS Survey} (\textit{LEnsing and Galaxy Growth: Observing Substructures}), aimed at studying the physics of clumpy star formation and quenching in eight lensed galaxies at $z\sim2$--4. LEGGOS combines multiple Cycle 2 \jwst GO programs (GO 4125, GO 3843) and Cycle 1 archival data, and utilizes strong gravitational lensing with NIRCam imaging and NIRSpec integral-field spectroscopy. LEGGOS targets UV-bright, highly magnified systems to resolve $\sim$10--200 pc regions in both rest-frame optical continuum and nebular emission. This overview paper describes the survey design, data reduction and calibration strategy, and science-quality data products, and highlights early examples demonstrating how spectroscopy breaks key degeneracies inherent to photometry-only clump studies, including identifying recent quenching in previously-thought UV star forming galaxies. We introduce a uniform analysis framework that jointly models lensing reconstruction, multi-band photometry, and integral field spectroscopy to disentangle multiple stellar populations within individual clumps and their surrounding diffuse regions. Using maps of Balmer recombination lines and key emission line diagnostic ratios, we connect star formation histories, dust attenuation, and nebular conditions on sub-kpc scales -- LEGGOS galaxies range from uniform metallicities across the whole galaxy, to having higher clump metallicities and harder ionization conditions relative to diffuse regions. The full survey dataset, with simultaneous flux and morphology constraints on clumpy source-plane regions, and a flexible spectrophotometric SPS modeling approach, provides a direct bridge between parsec-scale star formation physics and galaxy assembly at and beyond cosmic noon, offering a robust and efficient means of resolving star formation in the first galaxies.

\end{abstract}

\keywords{gravitational lensing, JWST, star formation histories, line ratios, SED fitting, starburst galaxies, post-starburst galaxies, spatially resolved observations}


\section{Introduction} \label{sec:intro}



Strong gravitational lensing transforms typically small, faint, marginally-resolved galaxies into bright highly-magnified arcs. The most magnified of these systems enable exquisite studies of internal galaxy morphologies down to scales of only tens of parsecs even in the most distant galaxies (e.g., \citealp[]{Bayliss_2014,Livermore_2015, johnson2017apjl_1110paperIII, Cornachione_2018, Rivera-Thorsen_2019, Ivison_2020, Florian_2020, welch2023}). Such lensed sources are exceedingly rare, and are generally found by searching many thousands of square degrees of ground-based survey imaging (e.g., \citealp{Allam_2007, Belokurov_2007,Koester_2010,Zang_2023}) or by dedicated imaging programs targeting massive galaxy clusters not previously explored (\citealp{Bleem_2015,Bleem_2020,Coe_2019}). Conversely, deep observations of known strong lenses with large lensing cross-sections (c.f.\ the \textit{Hubble} Frontier Fields; \citealp{Lotz_2017}) enables the discovery of many more lensed galaxies, but these systems are magnitudes fainter than the rare systems found in wide searches.

Studies of the brightest lensed galaxies are complementary to those of faint lensed and unlensed field galaxies -- the former offer enhanced source-plane spatial resolution and the photon flux for high-resolution spectroscopy, while the latter provide the statistical samples that rare, bright lenses cannot. The brightest strongly-lensed galaxies are the observational signposts with which we navigate the more abundant but much fainter deep-field samples. To date, searches of the Sloan Digital Sky Survey (SDSS; \citealp{Ahumada_2020})  have identified hundreds of bright lensed sources at $z$<3, the brightest of which typically have integrated g$_{AB}$ magnitudes of $\sim$20 \citep[e.g.,][]{Kubo_2010,Bayliss_2011,Stark_2013}.

This success is due to the depth and filter selection of the SDSS, sufficient to find the bright blue arcs that are the hallmark of UV-bright lensed galaxies at these redshifts. Many of these lensed targets have been followed up with HST imaging, revealing details of star-formation and structure on spatial scales down to tens of parsecs (e.g., \citealp{johnson2017apjl_1110paperIII,Cornachione_2018}). The brightest of these lensed galaxies --- a few of which predate the discoveries from the SDSS --- have been followed up with high-quality spectroscopy (e.g., \citealp{Pettini_2000,James_2014,Rigby_2018a,Rigby_2018b}) that exceed in both spectral resolution and signal-to-noise than what is available for stacks of dozens or even hundreds of field galaxies (\citealp{Shapley_2003, Steidel_2016}). DECaLS, DELVE and HSC have generated further robust datasets that will enable the discovery of exquisite lensed arcs at $z>5$ \citep{Khullar2021, welch2023, Klein2024, hutchison2026}.

The building blocks of these galaxies, or any such galaxy, are star-forming ``clumps'', which in this context is a rather colloquial term for regions within a lensed arc with high surface brightness, and a concentrated, symmetric light profile. These are either star clusters, regions highlighting giant molecular clouds, or even globular clusters \citep{johnson2017apjl_1110paperIII, Mowla2022, pascale2023,claeyssens2023,claeyssens2025,glimpse2026,adamo2024,adamo2025, Whitaker2026}. Clump formation and evolution are governed in part by physical processes acting at the galaxy scale and beyond. At no epoch is this relationship more apparent than at cosmic noon, $z$$\sim$1--3, when galaxies formed their stars (and clumps) the most rapidly, and when clumps dominated galactic morphologies and experienced more extreme physical conditions (e.g., low metallicities, large sizes, and more intense ionizing radiation) than in the local universe. Since early Universe galaxies have now been found to evolve in distinctly different ways than at z$\sim$0, this begs the question: what is the astrophysics that is leading to clumpier SF at cosmic noon?

JWST's first science data \citep{Pontoppidan22}, on lensing cluster SMACS\,J0723.3$-$7327, demonstrated the combined power of JWST and gravitational lensing to measure clump demographics (and in turn, highlight physics) ---age, stellar mass, size, star formation rate (SFR), dust content, metallicity, and ionization parameter (the number of ionizing photons per hydrogen nucleus)---at the smallest scales.  Lensing combined with NIRCam resolution revealed dozens of clumps -- star clusters, globular clusters, giant molecular clouds, etc. -- within lensed sources in SMACS\,J0723.3$-$7327. NIRCam multi-band photometry enabled precise age--dating for these clusters through spectral energy distribution (SED) fitting \citep{claeyssens2023,Mowla2022,forbes2023,vanzella2022b}.  Such analysis is out of reach with \textit{Hubble} Space Telescope (HST) data due to lack of sensitivity/access to age diagnostics without being able to probe the features of older stellar populations. The JWST data reveal that some clusters are young; some are already up to 1 Gyr old---globular clusters captured in adolescence (e.g., \citealt{Whitaker2026})!  NIRSpec spectroscopy of distant galaxies in the same field revealed the physical properties of ongoing star formation using the familiar rest-frame optical diagnostics (e.g., \citealt{Osterbrock1992}), which JWST makes accessible at almost any redshift.

\begin{figure*}
\centering
    \includegraphics[width=0.95\textwidth]{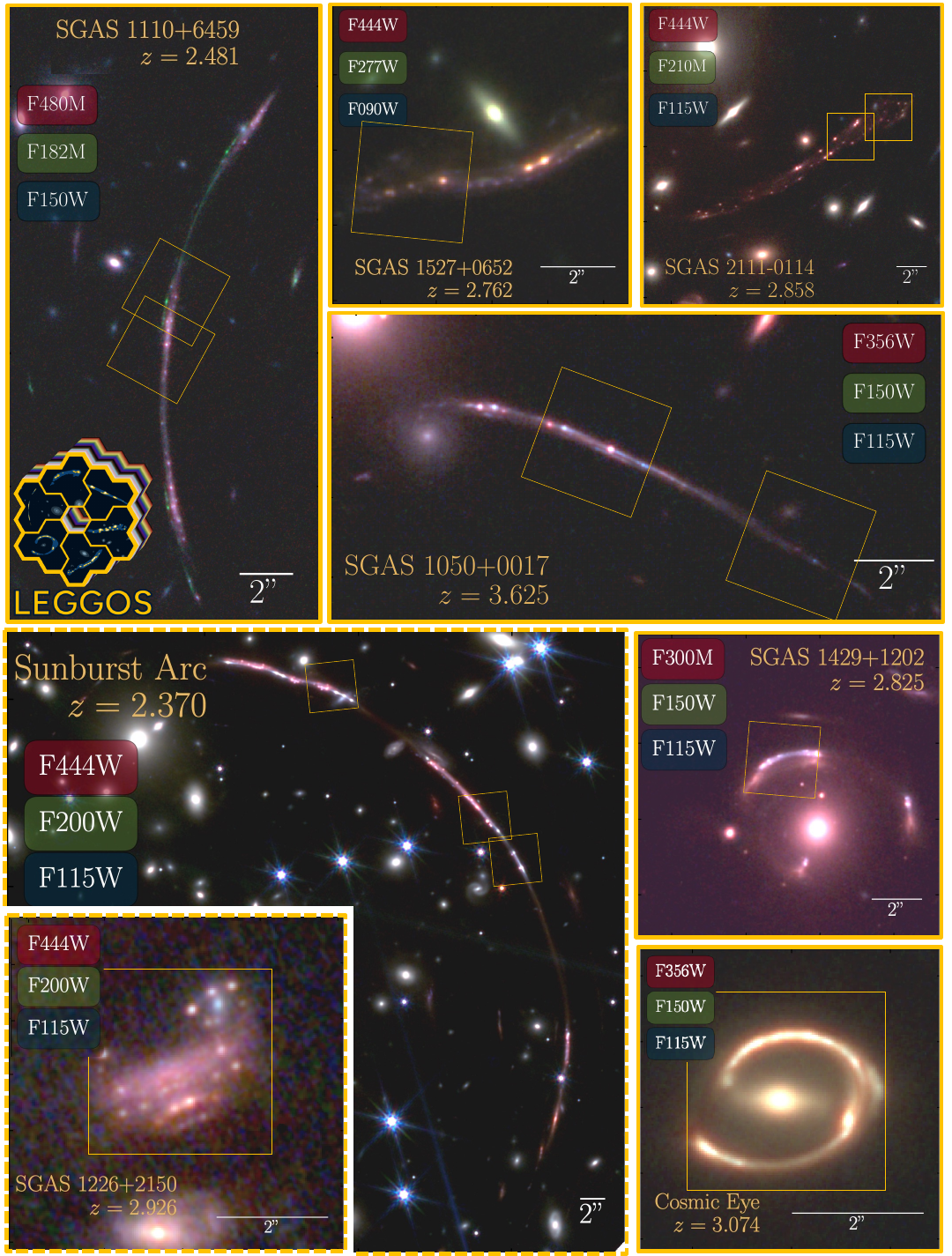}
    \vspace{-2mm}
    \caption{RGB mosaics of $JWST$ LEGGOS Survey targets. Each subplot notes the filters that make up the synthetic RGB image, the target name and spectroscopic redshift. Tens of clumpy regions are detected in each "primary" lensed arc, with half being sampled spectroscopically with NIRSpec/IFS (FOV in gold squares). (Bottom Right) The survey logo, highlighting the six new observed targets and the IFS component of the survey -- one of the hallmarks of this program. Also shown are the two archival targets Sunburst Arc and SGAS1226 (bottom left, dotted squares).} 
    \label{fig:mosaic}
\end{figure*}

Here, we present the JWST LEGGOS Survey, a spectrophotometric observation program of eight bright gravitationally lensed galaxies at $z=2-4$ (see Figure \ref{fig:mosaic}), probing 100s of clumps, the fundamental building blocks of these systems. Our survey relies on multiwavelength spectrophotometric observations, which simultaneously constrain star formation histories of stellar populations, nebular conditions, morphologies, light profiles and lens models of objects in these lensing fields, leading to a systematic (and relatively unbiased) picture of regions at the smallest physical scales beyond cosmic noon.

This paper is structured as follows. Section \ref{sec:survey} describes the motivation, science goals, target selection and deliverables of the survey, Section \ref{sec:data} unpacks the observations in this program, and Section \ref{sec:methods} lays out the various methodologies used to achieve our science goals. In Section \ref{sec:discussion}, we share some initial results and outcomes of the survey, and summarize our findings and next steps in Section \ref{sec:summary}.

All reported magnitudes are calibrated to the AB system. The fiducial cosmology models used in SED fitting assume a standard flat cold dark matter universe with a cosmological constant $(\Lambda$CDM), corresponding to WMAP9 observations (\citealp{Hinshaw2013}) (while lens models were computed with h=0.7, $\Omega_M=0.3$, and $\Omega_\Lambda=0.7$; see \citealt{bayliss2015_lens} for details on the specific impact of cosmological parameters on various relevant lens model outputs). For inferred parameters with uncertainties, we report 16th, 50th and 84th percentile values, unless otherwise specified. In this work, equivalent width is reported as $W_{line}<0$ for absorption features.


\section{The \JWST LEGGOS Survey} \label{sec:survey}

\subsection{LEGGOS is a Unique Survey, with Robust Methodologies and New Perspectives}

With LEGGOS, we have designed a \textbf{``survey of clumps"}, that combines JWST's spectroscopic and imaging capabilities, the magnifying power of gravitational lensing, along with archival HST data, to determine demographics for $>100$ individual SF/quenched clumpy regions (and a total of $>300$ clump images due to lensing multiplicity) in a sample of 8 strong gravitationally lensed galaxies at $z=2.3-3.7$, an epoch of the Universe (spanning 1.2 Gyr) where the rate of change of the cosmic star formation rate density is at its highest \citep{Madau2014}. With LEGGOS, we are attempting to chart the assembly of stellar mass within these galaxies, clump by clump (whether they be star clusters, globular clusters, giant molecular clouds, or bright lensed knots), and determining the detailed physical conditions within these small and dense structures. Our simultaneous treatment of flux profiles and morphology with forward modeling, and stellar population synthesis fitting is one of the hallmarks of this program. The series of studies based on the JWST LEGGOS Survey will address questions encompassed by two themes.

First, LEGGOS focuses on the construction of galaxies. Clump ages and their spatial distributions within galaxies (age gradients) teach us about the mechanisms of bulge formation and growth \citep{noguchi1999, elmegreen2008, bournaud2014, oklopcic2017, sachdeva2017}, as well as the dominant modes of mass assembly (hot/cold accretion versus major/minor mergers) (e.g., \citealp{dekel2009, hopkins2009_2,keres2009, oser2010,nakazato2024,sok2026}). Resolving critical scaling relations -- for example, the SFR vs stellar mass, and the mass--metallicity relations -- at the clump scale will reveal the origin of these galaxy-scale scaling relations at this epoch in the Universe.

Secondly, LEGGOS investigates the nature of small-scale structures, like molecular clouds and HII regions.
Clump stellar mass functions below 10$^{7}$M$_{\odot}$ directly probe the origins of clumpy structures \citep{knutas2025,faustinovieira2026,lapeer2026}. Relationships between clump sizes down to the tens of parsec scales and other demographics (e.g., SFR, stellar mass, and ionization conditions) will shed light on various aspects of internal physics of star-forming regions and their interactions with their host galaxies at large.

\begin{figure}
    \centering
    \includegraphics[width=0.5\textwidth]{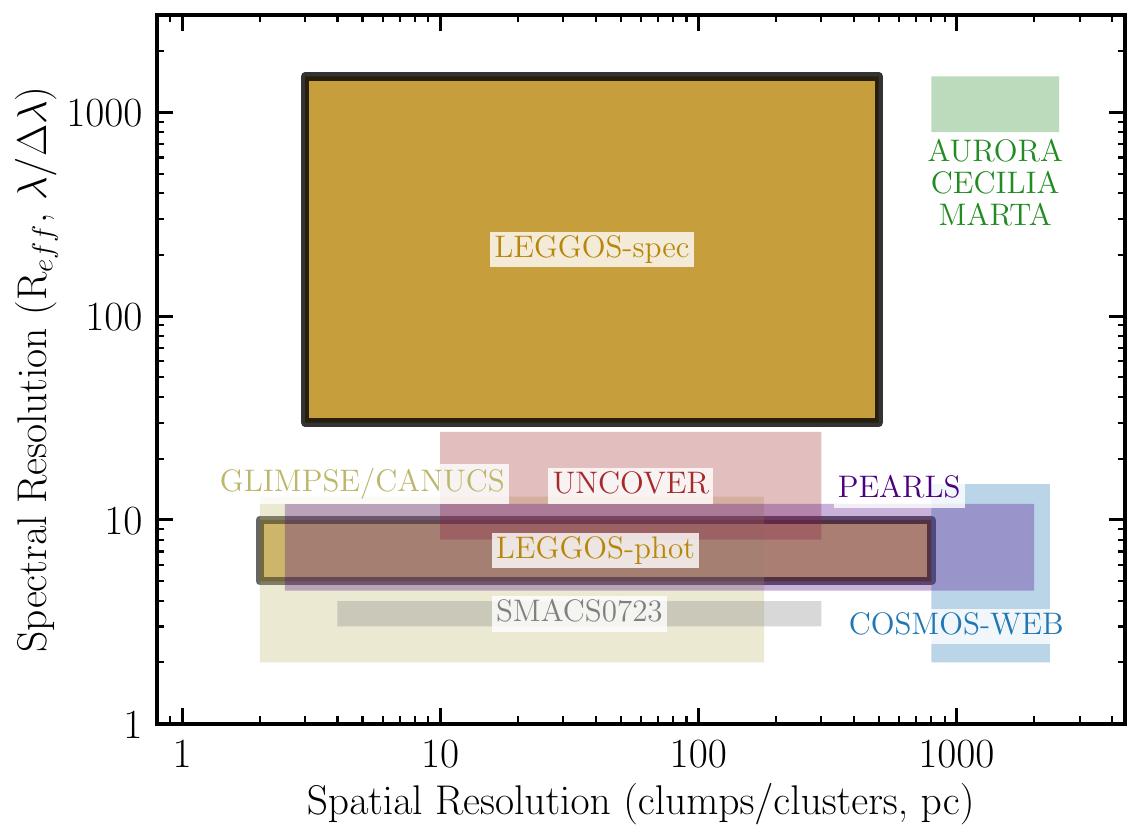}
    \caption{Effective Spectral vs Spatial Resolution for clump/star cluster-like observations in \JWST surveys.}
    \label{fig:survey_parameter_space}
\end{figure}

LEGGOS builds on the work of extant high-redshift lensed galaxy surveys, by specifically targeting the smallest highly-magnified substructures of galaxies beyond cosmic noon. Surveys like AURORA (GO 1914 PIs: Shapley \& Sanders; \citealp{Sanders2025}) and CECILIA (GO 2593 PI: Strom; \citealp{strom2023}) are studying cosmic noon ($z=2-4$) galaxies globally with high-resolution spectroscopy to chart nebular conditions in detail (e.g., metallicity studies, empirical calibrations for HII regions, ionization conditions, etc.), while surveys like COSMOS-Web (GO 1727  PIs: Kartaltepe \& Casey; \citealp{casey2023}), CANUCS  (GTO 1208 PI: Willott ; \citealp{willott2022}) and PEARLS (\citealp{windhorst2023}) are unveiling a discovery space of objects with clumpy starburst episodes, as well as post-starburst galaxy-like star formation histories. GLIMPSE (GO 3293  PIs: Atek \& Chisholm; \citealp{atek2025}) and UNCOVER (GO 2561  PIs: Labb\'e \& Bezanson; \citealp{bezanson2024}) are covering a niche parameter space by including the effect of a gravitational lensing boost to identify and characterize low-mass galaxies in the early Universe that are serendipitously found behind lensing clusters. LEGGOS is complementary to these efforts both in terms of accessing spatial resolution and inferring critical properties via low- and medium-resolution spectroscopy (see Figure \ref{fig:survey_parameter_space}). The LEGGOS target sample comprises well studied rest-frame UV bright objects, and the objective is to conduct targeted spectroscopic studies in a family of clumps in (previously thought) star-forming galaxies. 

Some of our unique features include:

\begin{enumerate}
    \item An emphasis on simultaneous spectrophotometric measurements and modeling of stellar populations and nebular line conditions with low and medium-resolution integral field spectrographs (JWST NIRSpec/IFS). These directly and robustly constrain age, dust attenuation and metallicities in these localized regions. 
    \item Non-parametric star formation history modeling with Bayesian SED Fitting (\texttt{Prospector}), with no underlying assumptions about stellar populations and parameterization of SF episodes in advance, nor assuming any connection between nebular and stellar continuum regions (unless the data suggests otherwise).
    \item An emphasis on the simultaneous characterization of clump morphology and fluxes in the source plane (not image plane), where both the imaging and ``lensing" point spread function are constrained (PSF, the instrument and lensing potential response to a given source plane pixel, or flux density from a clump).
\end{enumerate}

\subsection{Science Goals and Outputs}

The LEGGOS Survey has a multitude of goals, spanning its extensive breadth of data quality (observations with high signal-to-noise ratio; S/N), wavelength range, and parameter space of properties within the sample galaxies. LEGGOS is designed to definitively address four interconnected science themes:

\begin{enumerate}

\item \textbf{Mass assembly in gravitationally lensed galaxies:} A central open question in galaxy evolution is whether clumpy galaxies are built primarily through gas accretion or through mergers. Closely related is whether clumps themselves play a major role in building bulges and dense cores within star-forming galaxies --- a question LEGGOS is uniquely positioned to answer given its resolved, high-S/N spectrophotometric coverage.

\item \textbf{Stellar populations, physical conditions, and chemical conditions of clumpy star-forming galaxies:} Resolved star formation histories encode the cumulative record of how galaxies have assembled their mass; LEGGOS will use these histories to understand the overall mass build-up in galaxies at cosmic noon. Alongside this, the survey will map how dust is distributed relative to stellar emission in clumpy star-forming systems --- a key input for accurate attenuation corrections and physical interpretation of the observed emission.

\item \textbf{Critical properties of clumpy regions within galaxies:} The survey will characterize the distribution of clump sizes across the LEGGOS sample, and test whether clumps at cosmic noon follow the size--mass relation established for Milky Way giant molecular clouds -- interrogating whether an evolution (or absence thereof) in the sizes and stellar masses of clumps from local to high-redshift systems can be identified. LEGGOS will also determine whether small (<$100$ pc) clumps follow the same SFR--size relation as their larger counterparts at $2 \leq z \leq 4$. In addition, the survey will test whether the extreme values of ionization-sensitive line ratios, O32 and [O III]/H$\beta$ observed at z$\sim$3 are genuinely caused by extreme star formation rates in the early Universe, or whether other physical mechanisms are at play.

\item \textbf{Star-forming clumps and their contribution to cosmic reionization:} LEGGOS will interrogate the role individual star-forming clumps play in re-ionizing the universe. The combination of gravitational lensing magnification and JWST's sensitivity makes LEGGOS one of the few surveys capable of resolving the substructure relevant to this question.

\end{enumerate}





The following are the data products and deliverables that will be made public once the analyses have completed -- observations of the complete lensed galaxy and SF clump dataset, with its spatially resolved spectroscopy and panchromatic photometry, has immense legacy value for the field of galaxy evolution, especially at the smallest spatial scales at the highest redshifts:

\begin{enumerate}
    \item Level 2 and 3 \JWST/NIRCam imaging, and clump photometric catalogs (from custom reductions and measurements).
    \item \JWST/NIRSpec/IFS spectra maps, and 1D spectra of integrated lensed arcs and SF clumps.
    \item HST+JWST-based Gravitational Lens Models. 
    \item Spectral energy distribution (SED) posterior distributions for stellar population synthesis models at the spatial scales of clumps, as well as for integrated light across each galaxy (using outputs of Bayesian SED fitting frameworks for stellar population synthesis -- SPS -- modeling like \texttt{Prospector} \citep{Johnson2017, Johnson2021a} and \texttt{piXedfit} \citep{uf2021}.
    \item Spatial maps of emission line strengths and ratios sampling key diagnostics of nebular regions across the lensed galaxy sample.
\end{enumerate}

\subsection{Target selection}

The parent catalogs of bright lensed galaxies for this program are \cite{sharon2020} and \cite{Rigby_2018a}. From a set of 40 unique fields containing lensed galaxies in these studies, we select sources that:

\begin{enumerate}
    \item \textbf{are gravitationally lensed.}  Even with the sharper-than-expected spatial resolution of JWST, lensing is still needed to probe spatial scales $<100$ parsecs.  By contrast, JWST's diffraction--limited resolution is 300 pc for distant field galaxies (using the diffraction limit of 1.1 $\mu$m from \citealt{jwst2023}). The median magnification of clumps in our sample is 15, corresponding to an intrinsic size of $<20$ pc in the source plane.

    \item \textbf{are highly magnified.}  We select lensed galaxies that have been magnified by factors of $\geq 10$, giving us at least a 10$\times$ boost in flux (or $\sim3-5\times$ zoomed-in view of clump sizes). For comparison, the median magnification of the 76 lensed galaxies studied by GLASS \citep{glass2022} is 2.7; only four sources are magnified by $\geq 10$. GLIMPSE -- AbellS1063 -- galaxies studied photometrically in \cite{glimpse2026} have a median magnification of $\sim10$. The sources in our sample are also much more highly magnified than typical sources in CLASH \citep{xu2016} \footnote{Xu et al.\ 2016 reports length-to-width ratio ($\ell$/w), a proxy for magnification, with typical values of  5--10.  In contrast, one of our least extended arcs, SGAS1429+1202, has $\ell$/w $>$ 12.} or the Frontier Fields ($\overline{\mu} \sim 2-5$; \citealt{johnson2014}).

    \item \textbf{are clumpy.} In order to build up a large sample of clumps without incurring the significant overheads required to observe a large number of lensed galaxies with JWST, we select galaxies that contain many clumps. The number of clumps per source ranges from the 5 unique clumps that were clearly identified (and each multiply-imaged) in HST data at the time of selection, to the 2-3 dozen clumps in each of SGAS1110, SGAS1527, and SGAS2111.

    \item \textbf{have extensive HST archival data and HST--based lens models:}  HST UVIS is more sensitive to young, massive stars than JWST is, which is important for constraining massive star ages and UV slopes. Moreover, each source has an existing published high-quality lens model \citep{Smail07discoverypaper, sharon2022, Pignataro2021sunburst, Diego2022sunburst, Sharon2022sunburst,solhaug2025}. High-quality lens models enable forward--modeling techniques like those in \cite{Johnson2017_1110paperI,johnson2017apjl_1110paperIII} to precisely measure clump fluxes and sizes, and translate from apparent image plane values to the intrinsic source plane properties.

    \item \textbf{have redshifts in the range 2.3--3.8.}  This redshift range minimizes the required observing time by placing both H$\beta$ and H$\alpha$ in a single grating, G235M. At medium spectral resolution,these lines are resolved from adjacent lines, to ensure high-quality SFRs ($\sim10-15$\% precision expected) and extinctions (within $\sim$0.1 magnitude in A$_{V}$ expected). These redshifts also sample the blended [OII] 3727/3729 \AA\AA\ in the prism, enabling the use of a number of strong emission line diagnostics of clump physical conditions (e.g., O32, R23, Balmer decrement, N2H$\alpha$; see \citealt{welch2024} and \citealt{hutchison2026} for examples). The exception to this redshift range is the Sunburst Arc, where each of these lines is observed with the NIRSpec IFU at sufficient spectral resolution as part of Cycle 1 GO 2555 (PI: Rivera-Thorsen; \citealt{riverathorsen2024, riverathorsen2025arxiv, choe2025, welch2025}).

    \item \textbf{have high-quality rest-frame UV spectroscopy.}  For 7 objects, downloadable calibrated spectra were published as part of the MEGaSaURA atlas \citep{Rigby_2018a}.  SDSS\,J1110$+$6459 has comparable published rest-UV spectroscopy from the MMT blue channel spectrograph and GMOS  \citep{Johnson2017_1110paperI}.  Rest-frame UV spectra for the Sunburst Arc were collected from Magellan (e.g., a subset are reported in \citealt{sunburst2017,Rivera-Thorsen_2019, mainali2022, owens2024}).  These rest-frame UV spectra, typically available for either the integrated light of a whole galaxy or for a single bright clump or two in a given arc, will be a critical comparison sample vis-a-vis rest-frame far-UV properties inferred from JWST observations. Pre-JWST observations are described and tabulated in Appendix C.
    
\end{enumerate}

Only a handful of galaxies reside in this intersecting parameter space --- this resulting sample of 8 lensed arcs being the vast majority of these --- because they were selected from surveys covering large fractions of the sky, and are among the best laboratories that will \textit{ever} be found at this epoch of the Universe. By selecting arcs with visible clumps in HST, the sample has $>$100 individual clumps (and nearly 300 clump images due to lensing multiplicity).

We expect that for a clump with a stellar mass of 10$^7$ M$_\odot$ magnified 10$\times$, we measure ages within 5--30\%, and $A_v$ within 0.1 mag. We expect strong line ratios sensitive to ionization parameter and metallicity (O32 and R23) to be well measured in LEGGOS, with precision limited only by reddening uncertainty (these quantities are all magnification-invariant). By selecting fields with the most robust lens models available --- anchored by HST observations (Sharon et al. 2020) --- the survey limits magnification uncertainties to $\sim$10--20\% (towards stellar mass and star formation rates estimates).

Our targets of interest, described in detail in the sections below, are referred to by many names, including scientific and colloquial titles that are too numerous to describe here. Unless required by the needs of specific citations and the description of past studies, from here on, we will refer to our targets of interest -- SDSS\,J1110+6459, SDSS\,J1527+0652, SDSS\,J1050+0017, SDSS\,J2111-0114, SDSS\,J1429+1202, SDSS\,J2135-0102, PSZ1\,G311.65-18.48 and SDSS\,J1226+2152 -- as the following, respectively: \textbf{SGAS1110, SGAS1527, SGAS1050, SGAS2111, SGAS1429, Cosmic Eye, Sunburst Arc, and SGAS1226} (see Table \ref{table:list} for more details).

\section{Data} \label{sec:data}

As part of GO programs 4125 (PIs: Florian, Khullar) and 3843 (PI: Bayliss), our new set of JWST observations have three prongs:

\begin{enumerate}
    \item NIRCam six-to-eight band photometry to cover the rest-frame optical continuum from young and old stellar populations,
    \item NIRSpec IFS -- integral field spectroscopy -- prism, to sample the stellar continuum across a unique image of each lensed arc at low resolution (R $\sim 30-120$), and
    \item  NIRSpec IFS G235M observations (R $\sim1000$) to obtain emission line fluxes for lines that are blended at the resolution of the prism.
\end{enumerate}

GO 3843 (PI: Bayliss) also observed \eleventen\ with the NIRSpec IFS G140H and G235H modes, making it a legacy dataset with the highest redshift IFS observations in all available modes (see Section \ref{sec:sgas1110_highres} for more details).  

These arcs are often much larger than the $3''\times3''$ field of view of the NIRSpec IFS. Fortunately, lensing multiplicity makes tiling over the full arc redundant. We instead planned our new observations to capture the maximum number of \textit{unique} clumps while minimizing the number of tiles. We observed three galaxies with one IFS pointing, and another three with two pointings each (see \autoref{fig:2111_lens} for an example of a 2-tile footprint in SDSS2111). Moreover, no target acquisition was needed in this program, since the IFU is much larger than the absolute pointing accuracy of JWST of 0.1" (1$\sigma$, \citealt{Rigby2023backgrounds}). 

Table \ref{table:list} lists the complete set of observations comprising the JWST LEGGOS Survey (pre-JWST observations are referenced in Appendix C) \footnote{All \JWST Cycle 2 data (i.e., non archival observations) used in this paper can be found in MAST: \dataset[https://doi.org/10.17909/70dh-0x38]{https://doi.org/10.17909/70dh-0x38} and \dataset[https://doi.org/10.17909/h90f-n539]{https://doi.org/10.17909/h90f-n539}.
}.


%
\begin{deluxetable*}{lcccccccc}[!ht]
\tablecaption{\label{table:nirspec}LEGGOS Targets and NIRSpec/IFS Observations}
\tablecolumns{8}
\tabletypesize{\normalsize}
\tablewidth{\textwidth}
\tablehead{
\colhead{Galaxy} & \colhead{PID} & \colhead{RA} & \colhead{Dec} & \colhead{\multirow{2}{*}{$z$}} & \colhead{\multirow{2}{*}{$z_{lens}$}} & \colhead{Observation}  & \colhead{$t_{exp}$} & \colhead{\multirow{2}{*}{Grating/Filter}}  \\ 
~\vspace{-7mm}\\ 
 \colhead{Name} & \colhead{} & \colhead{(deg)} & \colhead{(deg)} & & & \colhead{Number}  & \colhead{[sec]} & \colhead{}  \\
\colhead{(1)} & \colhead{(2)} & \colhead{(3)} & \colhead{(4)} & \colhead{(5)} & \colhead{(6)} & \colhead{(7)} & \colhead{(8)} & \colhead{(9)} }
\startdata
    SGAS1050 & 4125 & 162.6641 & $+$0.2915 & 3.625 & 0.388 & 26 & 8870 &  G235M/F170LP \\
    & & & & & & 25 & 5952 & prism/clear \\
    Cosmic Eye & 4125 & 323.8029 & $-$1.0286 & 3.074 & 0.730 & 18 & 8870 & G235M/F170LP \\
    & & & & & & 27 & 5952 & prism/clear \\
    SGAS2111 & 4125 & 317.8280 & $-$1.2415 & 2.858 & 0.6363 & 16 & 8870 & G235M/F170LP \\
    & & & & & & 22 & 5952 & prism/clear \\
    SGAS1429 & 4125 & 217.4788 & $+$12.0439 & 2.825 & 0.531 & 14 & 8870 & G235M/F170LP \\
    & & & & & & 21 & 5952 & prism/clear \\
    SGAS1527 & 4125 & 231.9389 & $+$6.8720 & 2.762 & 0.392 & 12 & 8870 & G235M/F170LP \\
    & & & & & & 20 & 5952 & prism/clear \\
    SGAS1110 & 4125 & 167.5831 & $+$64.9978 & 2.481 & 0.659 & 10 & 8870 & G235M/F170LP \\
    & & & & & & 19 & 5952 & prism/clear \\
    & 3843 & & & & & 4 & 10737 & G140H/F100LP \\
    & & & & & & 2 & 7119 & G235H/F170LP\tablenotemark{$\dagger$} \\
    & & & & & & 5 & 7119 & G235H/F170LP\tablenotemark{$\dagger$} \\
\tableline
   SGAS1226 & 1355 & 186.7137 & $+$21.8722 & 2.926 & 0.436 & 1 & 4143 & G235H/F170LP \\
   Sunburst & 2555 & 237.5183 & $-$78.1833 & 2.370 & 0.443 & 1 & 5894 & G140H/F100LP \\
   & & & & & & 1 & 8870 & G235H/F170LP \\
\enddata 
\tablecomments{(1) Galaxy name, (2) Proposal ID, (3)--(4) NIRSpec IFU primary pointing coordinates (J2000), (5) Source redshift, (6) Lens redshift, (7) Observation number, (8) Exposure time per pointing -- note that SGAS1050, SGAS2111, SGAS1110, and the Sunburst arc include multiple pointings to cover the full extent of the arc, (9) Grating and filter combination used for the observation. }
\tablenotetext{\dagger}{Observations re-observed due to a mirror tilt event. Obs.\ 2 is the original, affected by the mirror tilt, and Obs.\ 5 is the re-observation.}
\label{table:list}
\end{deluxetable*}

\subsection{JWST Spectroscopy} \label{subsec:specredux}

Our observations with NIRSpec IFS \citep{boker2022} are designed to obtain the following signal-to-noise ratios: (a) Nebular line emission in NIRSpec G235M:  S/N $>$10 integrated over H$\beta$ and [O III]~5008, and (b) Stellar continua with NIRSpec PRISM: S/N$>$5 per resolution element, to conduct Bayesian SED fitting with stellar population synthesis modeling with data across rest-frame UV to near-IR. This enables us to characterize stellar mass, star formation histories (SFHs) and dust properties of the clumps. 

We reduce our IFS data using the JWST data reduction pipeline version 1.20.2 \citep{Bushouse25_jwstpipe1p20p2}, and calibration reference data set (CRDS) mapping \texttt{pmap1466}. 
We utilize the NSClean algorithm \citep{Rauscher24_NSClean} to remove correlated read noise from the detector images.                       
Additionally, prior to running the \texttt{spec3} pipeline, we flag and remove bad pixels using two methods. 
The first is a top-hat cut on pixel values, with cut thresholds based on visual inspection of the calibrated files (\texttt{cal.fits}). We identify the brightest emission lines in the calibrated files, and set the upper limit at roughly twice this value to ensure no bright target pixels are accidentally removed. 
We also manually remove pixels with data quality flags set to any of the following: \texttt{NO\_SAT\_CHECK}, \texttt{UNRELIABLE\_FLAT}, \texttt{TELEGRAPH}, \texttt{MSA\_FAILED\_OPEN}, or \texttt{OTHER\_BAD\_PIXEL}. 
Finally, we run the pipeline's built-in outlier detection step to remove additional outliers and bad pixels. 

The LEGGOS data were observed using an 8-point small cycling dither pattern. This dithering strategy was chosen to better reconstruct the NIRSpec PSF, which is undersampled by the standard 0\farcs1 detector pixels, in an attempt to study the small-scale structures within the target galaxies. 
In order to maximize the scientific return of this dataset, we therefore produce two output data cubes with different resolutions. The first cube has the standard 0\farcs1 spaxel (refering to a spectral pixel within the IFS cube) size. The second cube is drizzled to an output spaxel size of 0\farcs03. The undersampling of the detectors is known to cause quasi-periodic oscillations or ``wiggles" in the spectra of point-like sources, especially when working with small spatial scales \citep[i.e. individual spaxels,][]{Law23drizzle}. 
The additional dithers are designed to mitigate this artifact. 

After the NIRSpec pipeline is run, we apply the \texttt{baryon-sweep} post-processing tool to remove any hot pixel or cosmic ray artifacts missed by the pipeline's outlier detection process \citep{Hutchison24baryonsweep}. 
Combining this post-processing tool with the standard pipeline outlier rejection algorithm has been found to yield optimal results for NIRSpec IFS data \citep{Hutchison24baryonsweep}. 

We utilize custom background subtraction tools for the IFS data. For medium- and high-resolution spectra, we follow \cite{Rigby2025} and use the predicted backgrounds from the JWST Backgrounds Tool (JBT). The predicted backgrounds are then subtracted from each spaxel to produce a final background-subtracted data cube. 

\begin{figure}
    \centering
    \includegraphics[width=\linewidth]{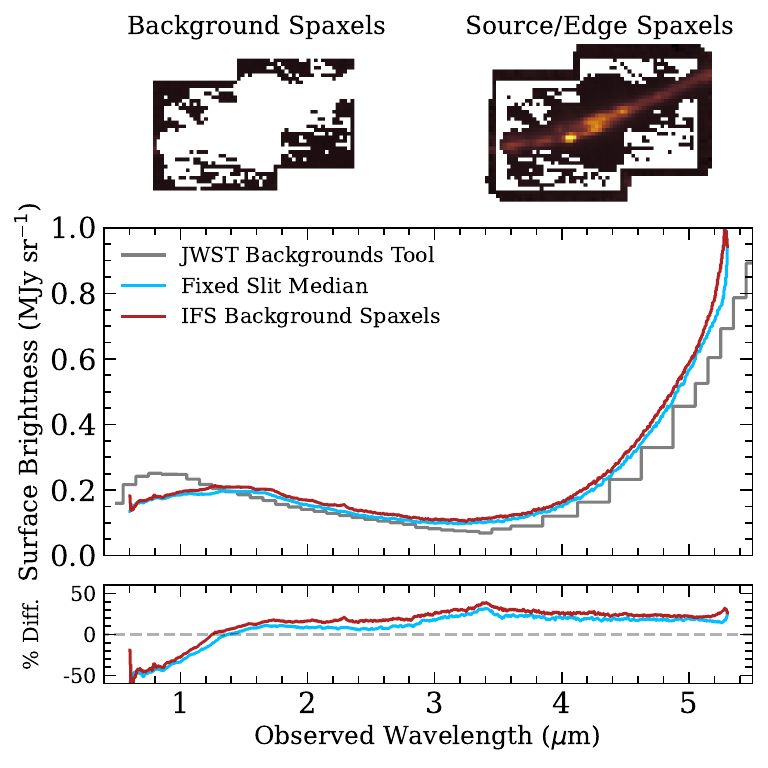}
    \caption{Backgrounds are measured locally in the NIRSpec/PRISM IFS data, as described in Section \ref{subsec:specredux}. The top panels show maps of \oiii\ for SGAS1110, with the top left panel including only the spaxels used to measure the local background, while the top right panel shows the source and edge spaxels which are not used in background estimation. The middle panel shows the extracted background spectrum from the IFS cube in red. As a comparison, we also show the predicted background from the JBT in grey, and the median background measured in empty fixed slits in blue (with percentage differences against JBT plotted in the bottom panel). The IFS and fixed slit backgrounds are consistent, indicating that our local IFS background is not contaminated by flux from the source. }
    \label{fig:prismbg}
\end{figure}

We take a more complex approach to subtract the background from PRISM observations. The JBT has a known issue below wavelengths of $\sim 1.2 \mu$m, causing the predicted background to be significantly elevated compared to measured backgrounds in multiple NIRSpec prism observations \citep[][see also Fig. \ref{fig:prismbg}]{Rigby2023backgrounds}. We therefore measure the background for each prism observation using off-target spaxels within the field of view. We define target spaxels using the S/N of the brightest lines in each cube (typically the [OIII]$\lambda5008$ emission line). Background spaxels are then those below the S/N cut. We further cut spaxels at the edge of the FoV from the background calculations, since these spaxels tend to be noisier due to the dithering pattern used. An example of this background spaxel selection for SGAS1110 is shown in the top panels of Figure \ref{fig:prismbg}. Once background spaxels have been identified, we combine the spectra into a single median background spectrum for each target. For the brightest targets, we then mask out bright emission lines from the background spectrum, and interpolate the background across these masked regions. This prevents faint wings of the brightest emission lines from contaminating our background measurements. This median background is then subtracted from each spaxel in the data cube, yielding a final background-subtracted cube for the prism. 

As a consistency check, we compare our extracted backgrounds from the off-source spaxels of the IFS to both the predicted backgrounds from the JBT, and the backgrounds measured in empty fixed slits. 
Because the fixed slits are always open, they provide an independent way to measure the on-sky backgrounds for each target. We reduce the fixed slit data using the same pipeline version and reference files as for the IFS data, and manually change the \texttt{EXP\_TYPE} header keyword to \texttt{NRS\_FIXEDSLIT} to tell the pipeline to extract the fixed slit data. 
The background spectra for the IFS, JBT, and fixed slits are plotted in Figure \ref{fig:prismbg} for SGAS1110. For clarity, only the median of the empty fixed slits is shown. The IFS and fixed slit backgrounds are consistent within uncertainties, indicating that our local IFS background is not contaminated by flux from the source. 
The consistency between the edge spaxels of the IFS and the empty fixed slits emphasizes the utility of these blank regions for background subtraction, even for extended lensed arcs. These findings support previous work recommending that off-target background observations are not necessary with the IFS \citep{welch2023rnaas}, as either the off-source IFS spaxels or the empty fixed slits provide sufficient constraints on the local sky background. 

\subsubsection{GO 3843 / High-resolution IFS spectroscopy of SGAS1110} \label{sec:sgas1110_highres}

We observed SGAS1110 with the NIRSpec IFU in two high-resolution grating settings --- G140H/F100LP and G235H/F170LP --- as a part of program GO-3843. At the redshift of the source, these data cover rest-frame $\sim$0.3-0.9 $\mu$m. 

The integration times in the two gratings were determined by using the JWST ETC; we estimate the time to achieve minimal measurements of key lines in the faintest resolved clumps in SGAS1110 \citep[m$_{\rm AB} = 27.7$ in HST F606W;][]{Johnson2017_1110paperI} using the same 
UV-through-NIR model spectra of individual clumps described above. 
We aimed for detecting the total [O II] $\lambda$3727,3729 doublet emission in the faintest clumps at S/N $\geq3$ in G140H/F100LP data, and H-$\alpha$ at S/N $\geq$10 in the G235H/F170LP data.

The resulting observations consist of total integration 
times of 10737s in G140H/F100LP and 7119s in G235H/F170LP. Observations with G140H/F100LP and G235H/F170LP were split across 16 and 8 exposures, respectively, to keep individual 
exposures under 1000s to mitigate the impact of cosmic rays on the final data. Both gratings were observed with an 8-point CYCLING-SMALL dither pattern to enable final datacube construction with spaxel sizes significantly smaller than the native 0.1". 

\subsubsection{Archival NIRSpec spectroscopy for the Sunburst Arc and SGAS1226}

We utilize archival NIRSpec/IFS spectroscopy from GO 2555 (PI: Rivera-Thorsen) for the \sunburst, with the G140H and G235H gratings. Four pointings were observed with NIRSpec; three on-target pointings and one off-target for background subtraction. Each of the NIRSpec pointings was observed in both the F100L/G140H and F170L/G235H settings, together covering the entire rest-frame optical range at
$2900 < \lambda < 9700 \AA$. The NIRSpec pointings were designed such that P1
covers the largest possible part of the galaxy in one pointing; P2 is centered around the peculiar clump in its center, named ``Tr'' in \cite{vanzella2020a}, or ``Godzilla'' by \cite{Diego2022sunburst}; and P3 is placed to cover two strongly magnified images of a Lyman-continuum emitting (LCE) cluster. Since
P1 lensing shear is nearly perpendicular to that of P2 and P3 \citep{Sharon2022sunburst}, this choice of pointings optimizes the amount of spatial information that can be recovered.

We also incorporate NIRSpec/IFS spectroscopy of \twelvetwentysix\ from JWST/ERS-TEMPLATES (PIs: Rigby, Vieira), with the G235H/F170LP grating (as well as a dedicated background pointing).

\subsection{Archival MIRI spectroscopy for \twelvetwentysix}

TEMPLATES \citep{Rigby2025} also contains archival MIRI/MRS spectroscopy in Setting C for SGAS1226, which we utilize in our broader dataset. We direct the reader to \cite{Rigby2025} for further details on data reduction, data products and analyses using MIRI observations.


\subsection{JWST/NIRCam Imaging}

The NIRCam imaging \citep{rieke2023} in GO 4125 is designed to span rest-frame $\sim$0.13--1.3 $\mu$m for all primary lensed arcs, and wherever possible, the filter strategy is implemented to sample stellar continuum while avoiding key emission features, though in certain cases, when medium band filters were particularly well-suited to capturing an important emission lines, they were used for that (e.g., F182M captures [OIII] for SGAS1110).  The integration times provide signal-to-noise (S/N) $\geq$10 measurements of the faintest resolved clumps (estimated via HST imaging and JWST exposure time calculator, or ETC), significantly motivated by the plan in ERS TEMPLATES \citep{Rigby2025}. 

We briefly describe the JWST data reduction process used to create science-ready imaging. We begin with the level 2A ``rate" files from MAST. We correct for 1/f noise and jumps between amplifiers in them by using a custom script. The resulting corrected Level 2A files are processed to level 2B using the standard pipeline. At this stage, shifts and rotations are applied using custom Python scripts in order to match the WCS solutions of the NIRCam data to existing archival HST data. The WCS- and 1/f-noise-corrected level 2B ``cal" files are then processed to create stage 3 science-ready mosaics aligned with the HST images. Throughout this process, we use jwst pipeline version 1.15.1 and the CRDS calibration reference data system pipeline mapping (CRDS, pmap) 1298. All images used in this paper were resampled to a $0\farcs03$ grid, using Gaussian kernels.

One of the most critical sources of noise in our imaging is 1/f amplifier noise, labeled flicker noise in STScI terminology \citep{Rauscher24_NSClean}, which manifests mostly as stripes of various lengths and amplitudes across the image, primarily in the horizontal direction (though some vertical striping is also seen, due to other effects). We attempt to remove this from the Level 2A ``rate'' files. The first step is to mask bright sources.  To do this, we use the level 3 ``i2d'' image created from the original level 2 files (that had not yet been corrected for 1/f noise).  These bright source masks are propagated back to each of the level 2A files and applied to them. We smooth out pedestal offsets between chips and amplifier regions by measuring any jump or drop across the transition between two amplifiers and correcting it and then measuring the sky in a grid of 256 pixels (while rejecting outlier pixels to minimize the impact of sources that may not have been fully masked in the first step) and correcting for any sky background that still remains.

With background matching on the largest scales measured and removed, we then filter the image horizontally and vertically on chip-wide scales on individual axis, using a $2048\times1$ pixel median filter in each of these directions, to remove large linear low frequency noise features. 1/f noise exists on all scales however, so filtering out yet smaller scale structured noise along the x axis (i.e., rows) of the individual images is still required. 

We use a ring median filter at several scales to construct  smoothed versions of each image, progressing from smaller to larger filtering scales iteratively. Each ring-smoothed image is subtracted from the input data (as filtered to that point) effectively removing most features with a near-circular profile (i.e., actual objects). The smallest ring filter used has an inner radius of 0.5 pixels (so that it doesn't include the signal from a given row) and an outer radius of 1.5 pixels. The largest are about an order of magnitude larger. A horizontal box median filter, at least 50 pixels wide (along the rows) and one or more pixels tall (along the columns, sized to match the inner radius of the corresponding ring median) is then applied to the residual image, isolating horizontal striping. The resulting median-smoothed residual image is then subtracted from the input data, and the resulting image is then passed to the next iteration of the filtering. Some residual noise at the smallest scales is unavoidably left in place because real sources at those scales prevent a clean separation of signal and noise. The scales of the filters used in this process have been carefully chosen, through some trial and error, to produce reasonably clean results, and the final de-striped level 2A images show clear improvement over the original files.  The de-striped level 2A files are processed through stages 2 and 3 of the standard NIRCam pipeline to create the final level 3 drizzled ``i2d" images, as described above, which we use for the analyses in this program.

In the F150W imaging for the field of \SDSSeleventen, inadequate cosmic ray flagging in the standard pipeline resulted in the type of NIRCam artifact commonly known as a ``snowball" in two locations near or overlapping the arc.  These were corrected by manually flagging the affected pixels in the data quality array of the cal files.  In several filters, for multiple targets, a number of bad pixels were not masked and resulted in diamond-patterned artifacts, effectively imprinting the dither pattern, in Level~3 mosaics. These, too, were fixed by flagging them in the data quality arrays outside of the standard pipeline.  For the cosmic eye, a diffraction spike from an exceptionally bright star outside of the field of view fell on top of the arc.  Because the feature aligns very closely with the pixel columns, we removed the spike by fitting each affected column with a polynomial in blank areas of the sky in the level 2A image near the arc, both above it and below it, and subtracted the predicted flux attributable to the diffraction spike. Those data were then processed through stage 2 of the standard pipeline, as we did for all other targets, to produce level 2B images. We re-ran Stage~3 of the pipeline, as described above, using the updated files for these targets; the regions containing the cluster and arcs in the final data products are unaffected by unmasked bad pixels, snowballs, and in the case of the Cosmic Eye---the diffraction spike.

For more details on the various steps and systematics involved in this process, please refer to the complete description of the custom pipeline provided in \cite{Rigby2025}.

All NIRCam imaging in GO-4125 and GO-3843 were taken using a 4-point sub-pixel dither pattern to support reconstruction of the exceptionally sharp NIRCam PSF in all bands. The total imaging integration times used for each filter and for each target are given in Table~\ref{tab:nircam}, along with the dither and readout patterns, as well as the groups per integrations and integrations per exposure.  Note that for SGAS1110, F480M observations in GO-4125 were paired with each of the short wavelength observations in F182M, F150W, and F200W in order to build up signal to noise, while broader wavelength coverage in the longer wavelength NIRCam channel was acquired as part of GO-3843.  This means F480M exposures were repeated three times, effectively tripling the total exposure time.

\begin{deluxetable*}{lcccccccccc}[!ht]
\tablecaption{LEGGOS NIRCam Observations}
\tablecolumns{6}
\tabletypesize{\footnotesize} 
\tablehead{
\colhead{Galaxy} & \colhead{PID} & \colhead{Filter(s)} & \colhead{Obs} & \colhead{Dither} & \colhead{Total} & \colhead{Readout} & \colhead{Groups/} & \colhead{Ints/} & \colhead{Repet-} & \colhead{Total Exp.}   \\ 
~\vspace{-7mm}\\ 
\colhead{Name} & & & \colhead{Number} & \colhead{Pattern} & \colhead{Dithers} & \colhead{Pattern} & \colhead{Int} & \colhead{Exp} & \colhead{itions} & \colhead{Time [sec]} \\
\colhead{(1)} & \colhead{(2)} & \colhead{(3)} & \colhead{(4)} & \colhead{(5)} & \colhead{(6)} & \colhead{(7)} & \colhead{(8)} & \colhead{(9)} & \colhead{(10)} & \colhead{(11)}}
\startdata
SGAS1050 & 4125 & F115W, F150W, & 24 & INTRAMODULEBOX & 4 & BRIGHT2 & 10 & 1 & 1 & 858.942 \\
& & F182M, F277W, & & & & & & & \\
& & F356W, F444W & & & & & & & \\
Cosmic Eye & 4125 & F115W, F150W, & 17 & INTRAMODULEBOX & 4 & BRIGHT2 & 10 & 1 & 1 & 858.942 \\
 & & F182M, F277W, & & & & & & & \\
  & & F356W, F444W & & & & & & & \\
SGAS2111 & 4125 & F115W, F150W, & 15 & INTRAMODULEBOX & 4 & BRIGHT2 & 10 & 1 & 1 & 858.942 \\
 & & F210M, F277W, & & & & & & & \\
 & & F300M, F444W & & & & & & & \\
SGAS1429 & 4125 & F115W, F150W, & 13 & INTRAMODULEBOX & 4 & BRIGHT2 & 10 & 1 & 1 & 858.942 \\
 & & F210M, F277W, & & & & & & & \\
  & & F300M, F444W & & & & & & & \\
SGAS1527 & 4125 & F090W, & 11 & INTRAMODULEBOX & 4 & BRIGHT2 & 10 & 1 & 1 & 858.942 \\
 & & F162M+F150W2, & & & & & & & \\
 & & F210M, F277W, & & & & & & & \\
  & & F300M, F444W & & & & & & & \\
SGAS1110 & 4125 & F182M, F150W, & 9 & INTRAMODULEBOX & 4 & BRIGHT2 & 10 & 1 & 1 & 858.942\\
 & & F200W & & & & & & & \\
 & & F480M & 9 & INTRAMODULEBOX & 4 & BRIGHT2 & 10 & 1 & 3 & 2576.82 \\
 & 3843 & F070W, F444W & 3 & INTRAMODULE & 4 & MEDIUM8 & 6 & 1 & 1 & 2490.931 \\
 & & F090W, F356W & 3 & INTRAMODULE & 4 & SHALLOW4 & 7 & 1 & 1 & 1460.201 \\
 & & F115W, F277W & 3 & INTRAMODULE & 4 & SHALLOW4 & 6 & 1 & 1 & 1245.465 \\
\tableline
    SGAS1226 & 1355 & F115W, F150W, & 3 & INTRASCA & 3 & BRIGHT1 & 5 & 1 & 1 & 289.893 \\
 & & F200W, F277W, & & & & & & & \\
 & & F356W, F444W & & & & & & & \\
    Sunburst & 2555 & F115W, F150W, & 3 & INTRAMODULEBOX & 8 & BRIGHT2 & 10 & 1 & 1 & 1717.883 \\
& & F200W, F277W, & & & & & & & \\
& & F356W, F444W & & & & & & & \\
\enddata 
\tablecomments{(1) Galaxy name, (2) ID of program that acquired the data, (3) NIRCam Filters, (4) observation Number within the program, (5) the dither pattern used for the imaging, (6) total dither positions, including both primary and subpixel positions if applicable---only data for the Sunburst Arc used 2 subpixel positions instead of 1, (7) the readout pattern employed, (8) Groups per intergration, (9) integrations per exposure, (10) number of times this imaging pattern was repeated---for SGAS1110, F480M was the long wavelength filter paired with each of three different short wavelength filters for simultaneous imaging, resulting in a total exposure time 3 times as long as for the short wavelength filters, (11) the total exposure time in seconds. The last two targets, offset from the others by a horizontal line, were drawn from archival data.}
\label{tab:nircam}
\end{deluxetable*}

\subsubsection{GO-3843 / SGAS1110}

In addition to the previous set of observations as part of GO 4125, we obtained NIRCam imaging of SGAS1110 in F070W, F090W, F115W, F277W, F356W and F444W filters. At the redshifts of the galaxy, these bands span rest-frame $\sim$0.13--1.3 $\mu$m, while avoiding bright nebular emission lines and hence providing diffraction limited measurements of the spatially resolved continuum emission. These data inform morphological measurements of the stellar populations across the rest-frame UV-through-NIR. 

The NIRcam integration times were computed to deliver signal-to-noise (S/N) 
$\geq$10 in measurements of the faintest resolved clumps in SGAS1110, which have m$_{\rm AB} = 27.7$ in HST F606W imaging \citep{Johnson2017_1110paperI}. The F606W magnitudes sample the rest-frame $\sim$1700\AA\ UV emission, which we use to scale a model spectrum spanning the rest-frame UV-through-NIR to predict the likely imaging depths required in each NIRCam band. We do this by modeling the complete UV-through-NIR spectrum of SGAS1110 generated from stellar continuum and nebular emission models with the Bayesian stellar population synthesis (SPS) modeling code \texttt{Prospector}. We use the best-fit SED model for SGAS1110 based on the available pre-JWST data (primarily multi-band HST imaging). The model spectrum is then normalized to each individual clump's F606W AB magnitude to generate input model spectrum for the JWST ETC. Similar to 4125, all NIRCam imaging for SGAS1110 in this program were taken using a 4-point sub-pixel dither pattern to support PSF reconstruction in all bands. The total imaging integration times used for SGAS1110 were 2490s in F070W/F444W, 1460s in F090W/F356W, and 1245s in F115W/F277W (determined, as in GO-4125, by the SNR requirements in LW filters). 

\subsubsection{Archival Imaging on the Sunburst Arc and SGAS1226}

The Sunburst Arc is the subject of a dedicated Cycle 1 GO program 2555 (PI: Rivera-Thorsen, NIRSpec IFS and NIRCam), with the full field being observed with NIRCam in the filters F115W, F150W, F200W, F277W, F356W, and F444W (see \citealt{riverathorsen2025arxiv} for more details).

Due to its redshift (and hence, the rest-frame wavelength coverage of its archival spectra), SGAS1226 was one of the targets observed via the JWST Early Release Science (ERS) program TEMPLATES (PI: Rigby, Vieira), with observations in the filters F115W, F150W, F200W, F277W, F356W, and F444W (see \citealt{Rigby2025} for details).  The bottom section of Table~\ref{tab:nircam} summarizes the observing strategy for both of these archival NIRCam targets.

\subsection{Archival MIRI Imaging}

We take note of archival MIRI imaging for SGAS1226 \citep[TEMPLATES ERS;][]{Rigby2025} and the Sunburst 
Arc (Cycle 3 GO 6353, PI: Pascale), the two LEGGOS galaxies that are archival within our program. 

TEMPLATES observed SGAS1226 with seven MIRI filters -- F560W, F770W, F1000W, F1280W, F1500W, F1800W, F2100W -- with a dither pattern as described in \cite{Rigby2025}, and a minimum S/N $>10$ per resolution element. GO 6353 observed the Sunburst Arc in all available MIRI filters -- F560W, F770W, F1000W, F1130W, F1280W, F1500W, F1800W, F2100W, and F2550W -- with a 10 point medium dither pattern. GO 6353 S/N considerations are calibrated to boost the signal from multiply imaged super star clusters candidates \citep{Rivera-Thorsen_2019, pascale2024}.

We will utilize these data points in our pan-chromatic SPS modeling wherever possible (See Sections \ref{sec:sps} and \ref{sec:sps_results} for more details).

\begin{figure*}[t!]
\centering
    \includegraphics[width=1.0\textwidth]{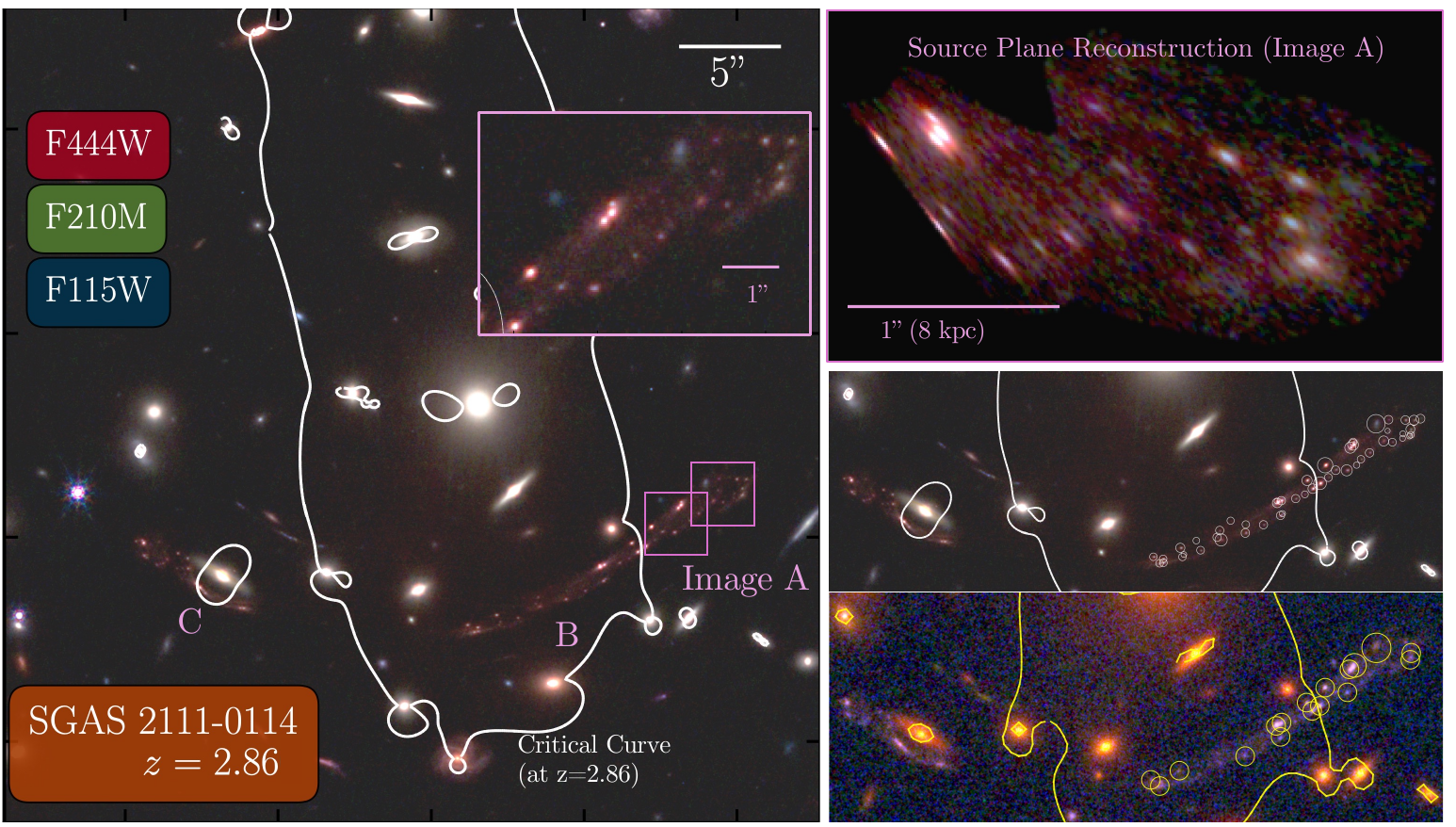}
    \caption{\textbf{(Left)} Gravitational Lens model for SGAS2111, with critical curve (locus of infinite magnification) at $z=2.86$. Also shown in purple squares (and the zoomed in top inset) is the primary image of the arc, sampled with NIRSpec/IFS. (Right)  \textbf{(Top right)} Source plane reconstruction of SGAS2111 using NIRCam imaging of Image A, the region which is spectroscopically sampled with the IFS, showing two distinct regions of clumpy and diffuse fluxes, potentially indicating a merger scenario. \textbf{(Middle and Bottom Right)} JWST mosaic for SGAS2111 centered between the three multiple images of the lensed arc, in direct contrast with HST imaging and old lens models for the same field. The number of identified clumps is much larger in JWST imaging relative to HST F160W/F814W/F390W (white vs yellow circles, respectively), which directly impacts the fidelity of the lens model, and demonstrates that the galaxy is much more comprehensively resolved in JWST imaging (and spectroscopy).} 
    \label{fig:2111_lens}
\end{figure*}

\subsection{Ancillary \textit{Hubble} and ground-based observations}

LEGGOS galaxies have been extensively studied with HST, across multiple GO programs as well as ground-based spectroscopy; we use archival observations wherever necessary to fold in rest-frame UV observations. We also assemble this information for the reader in Appendix C, describing the facility, type of observation and corresponding references. 

\subsection{Deep F555W imaging of SGAS1110 -- GO 3843}

As part of JWST GO~3843, we also obtained new HST WFC3/UVIS imaging of SGAS1110 in F555W, executed in HST Cycle 30 (HST-GO-17311; PI: Bayliss). These data consist of 5 orbits of WFC3/UVIS F555W imaging data that measures the rest-frame UV continuum emission from the faintest known clumps in SGAS1110 at S/N $\geq$10 (photometric uncertainties of $\sim \pm 10$\%), matching the precision of the NIRCam imaging data. F555W is centered at rest-frame 1590\AA\ for SGAS1110 while completely avoiding Lyman-$\alpha$, so that these data provide the bluest-possible rest-UV measurements of the stellar continuum emission in SGAS1110.

These data were reduced using the \texttt{astrodrizzle} function of \texttt{DrizzlePac} \citep{Fruchter2010,Hoffman2021}, employing a Gaussian kernel, exposure time weighting, and final drop size of 0.8.  WCS solutions matching these data to the NIRCam data and previous reductions of archival HST data were conducted using \texttt{tweakreg}.  The final data were drizzled onto a North-up grid with 0.03 arcsecond pixels, resulting in images aligned with the data from other filters (including NIRCam filters) in both WCS and pixel space.  All other archival HST for each target was reduced similarly.

\section{Methods} \label{sec:methods}
\subsection{Gravitational Lens Modeling}
 
Translating the observed properties of gravitationally lensed galaxies to their intrinsic values requires accounting for the effects of magnification and distortion. In particular, physical size and any property that is derived from flux measurements, scale with the magnification. Such properties include clump sizes, luminosities, SFRs, and stellar masses. Since the lensing is achromatic, measurements that rely on flux ratios are invariant to lensing magnification, e.g., observed color and emission line ratios, and subsequently inferred metallicities, mass-weighted ages, and specific star formation rates. Still, understanding these measurements in the context of their source morphology, size, and other physical properties, requires accounting for the lensing effect.

Gravitational lens modeling analyzes the strong lensing evidence---the occurrences of multiple images of the same background source---and results with a mass map of the foreground lens from which lensing outputs such as magnification and deflection maps can be derived. For its lens modeling analyses, the \LEGGOS\ survey employs the strong lens modeling algorithm \lenstool\ \citep{Jullo2007}, which uses a ``parametric'' approach, i.e., the lens plane is modeled as a combination of parametric mass halos. The software uses Markov Chain Monte Carlo (MCMC) sampling to explore the parameter space and find a lens solution that minimizes the scatter between model-predicted and observed locations of lensing features. For details of the lens modeling procedures, we refer the reader to \cite{sharon2020}.

By design, the \LEGGOS\ targets were selected for their highly magnified giant arcs, which occur due to a fortuitous position of the background galaxy relative to the foreground cluster. The result of this lensing selection is that some of the LEGGOS lenses -- even with JWST observations -- are not as rich in lensing constraints as other famous lensing fields which reveal numerous lensed background galaxies, including in JWST programs like UNCOVER/GLASS's Abell 2744 \citep{glass2022,bezanson2024}, GLIMPSE's AbellS1063 \citep{atek2025}, or massive galaxy clusters observed in the JWST-SLICE Survey \citep{cerny2025}. Nevertheless, the clumpy nature of the LEGGOS arcs means a high density of constraints and a tight sampling of the lensing potential by the arcs, which significantly reduces lensing-related uncertainties, as demonstrated in \cite{Abedi2026} in \eleventen.

All LEGGOS fields have been modeled prior to the JWST program, and our analysis builds upon these works. Four of our targets were modeled as part of the SGAS-HST program (HST Cycle 20 GO-13003, PI: Gladders; \citealt{sharon2020}): \SDSSeleventen, \SDSSfiftintwentyseven, \SDSStenfifty, and \SDSStwentyone. These models relied on the HST imaging and ancillary spectroscopy from various ground based observatories.  
The lens model for \SDSSeleventen\ was first published in \cite{Johnson2017_1110paperI} and subsequently used in \cite{rigby2017_1110paperII} and \cite{johnson2017apjl_1110paperIII}, to constrain the source size of star forming clumps in the source galaxy. 
In \SDSSfiftintwentyseven, the LEGGOS galaxy is primarily lensed by a nearby galaxy at $z=0.43$, but the lensing line of sight is complicated by the lensing potential of a foreground cluster of galaxies at $z=0.392$.  \cite{sharon2020} used an iterative approach to model the underconstrained cluster and concluded that while the lensing configuration is well understood, higher accuracy requires deeper space-based data. 
The HST-based model for \SDSStwentyone\ was limited by the lack of reliable constraints north of the cluster core, which increases uncertainties due to modeling degeneracies. 
We refer the reader to \cite{sharon2020} for a detailed description of these pre-JWST lens models.

The HST-based lens model of \SDSSfourteentwentynine, presented in \cite{catan2024}, is based on data from HST Cycle 25 GO-15378 (PI: Bayliss). 
The most recent HST-based model of \SDSStwelvetwentysix\ was published in \cite{sharon2022} as part of the first data release of the JWST ERS-TEMPLATES program \citep{Rigby2025}. 
Models for the \sunburst\ lens have been published in several works \citep{Pignataro2021sunburst,Sharon2022sunburst,Diego2022sunburst}; LEGGOS in particular has built upon the \lenstool\ model published in \cite{Sharon2022sunburst}. Finally, a model of the \cosmiceye\ galaxy-galaxy lens was published in \cite{dye2007cosmic} based on a single HST band (F606W) that was available at the time (Cycle 14 GO-10491; PI: Ebeling). The lensing potential in this line of sight is complicated by shear from the nearby foreground cluster, MACS2135. A model of the foreground cluster was published by \cite{zitrin2016} adding multiband HST data from programs GO-12884 and GO-12166 (PI: Ebeling), but the lens modeling analysis does not extend to the \cosmiceye.
This field  therefore requires a complete re-modeling to properly account for the lensing potential of the foreground cluster out to the \cosmiceye, building on the lensing constraints identified in previous works and the archival HST data that cover the cluster.

For reference, from the extant models, we find the median magnification of clumpy regions in our sample across all arcs is 15, with dozens of clumps magnified by factors of 20--100, and a few by much more -- leading us to probe regions ranging from 10-100 pc, primarily because of the lensing boost.

In this Survey Paper, we show an example of a LEGGOS-enabled lens model of one of the LEGGOS targets, \SDSStwentyone.  The LEGGOS JWST/NIRCam imaging reveals new candidate lensed sources north of the BGC, a region that was under-constrained previously. We also identified and mapped twenty clumps between multiple images of the lensed galaxy that form the giant arc, more than tripling the number of identified multiply-imaged clumps in our HST-based model \citep{sharon2020}. In addition, spectroscopic confirmation of clumps that fall within the NIRSpec IFS field further solidify some identifications. The added constraints improve the modeling uncertainties significantly, especially along the LEGGOS target. We find very similar improvements in \SDSSeleventen, which we present in a companion paper \citep{Abedi2026}: the addition of new constraints from JWST resulted in reduction of the magnification uncertainties along the arcs by a factor of $\sim2-8$ (to $<10\%$).

\begin{figure*}
\centering
    \includegraphics[width=1.0\textwidth]{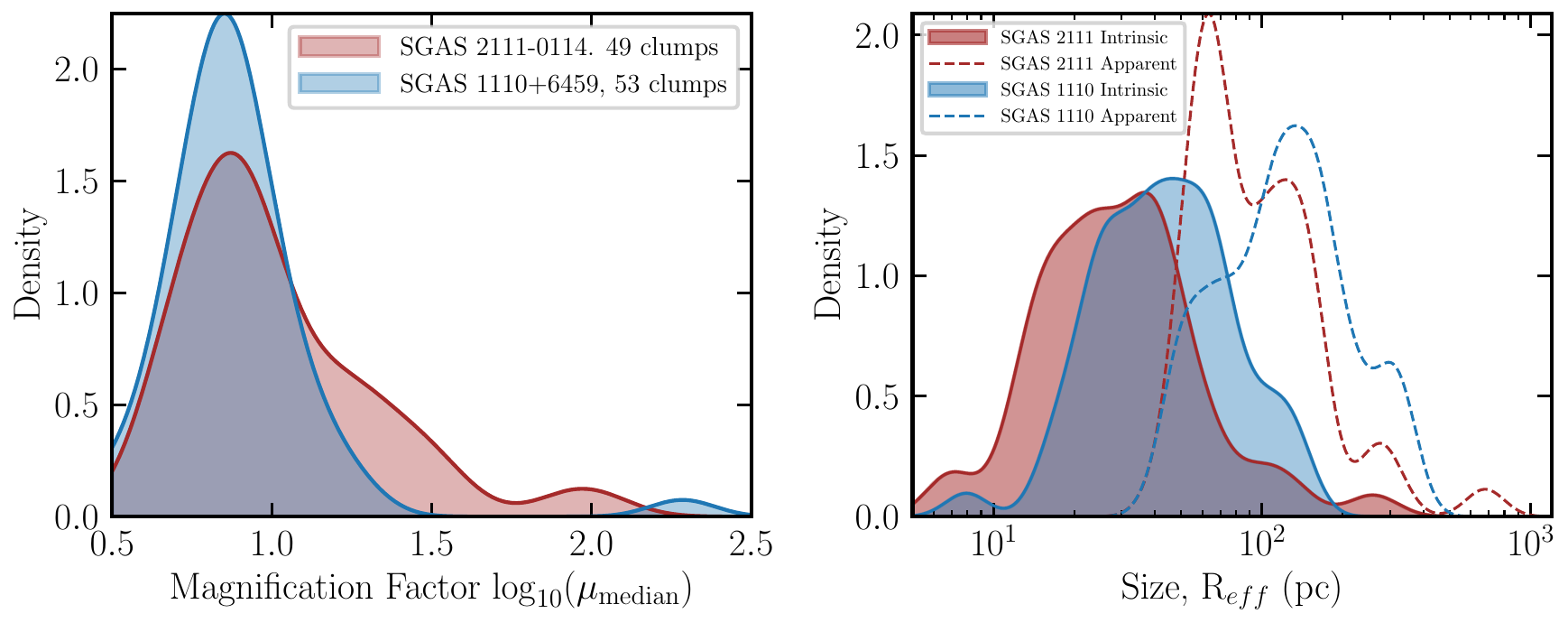}
    \caption{Probability distributions of magnification and intrinsic/source-plane sizes (R$_{eff}$) of clumps across galaxies SGAS2111 (red) and SGAS1110 (blue), estimated after Monte Carlo sampling the probability distributions of lens-model based magnification and effective radii in the lensing image plane. LEGGOS will sample tens of 10-200pc high surface brightness regions within the sample galaxies.} 
    \label{fig:mag_size_pdf}
\end{figure*}

In Figure \ref{fig:2111_lens} we show the best-fit lens model for \SDSStwentyone\, with the critical curves corresponding to the lensed arc redshift ($z=2.86$); we overlay the NIRSpec/IFS field-of-view and showcase (in the inset plot) the unique image of \SDSStwentyone\ being sampled with spectroscopy. The top-right panel of \autoref{fig:2111_lens} show the projection of the arc to the source plane using the best-fit lens model. The bottom-right panels show a comparison between the JWST- and HST-based imaging data and lens model, highlighting the increase in identified clumps.
\ref{fig:mag_size_pdf} shows the image and source plane sizes of clumps (once corrected for magnification) being sampled by our JWST imaging -- we probe SF (and quenching activity) in 20-200pc regions within this system.

LEGGOS analyses will also incorporate the impact of changing cosmological parameters ($\Omega_M$ and $\Omega_\Lambda$) on the magnification lensing deflections (mostly agnostic to $H_0$, and weaker dependence on $\Omega$); see \citep{bayliss2015_lens} for more details.

\subsection{Selecting and Modeling Clumps \label{subsec:selection}}

As a demonstration, in this work we use a multitude of approaches to measure the sizes and fluxes of clumps, with a ``forward-model'' of a source plane light distribution being the fiducial model in future LEGGOS studies. 

One of the most striking advantages of NIRCam is its sharp PSF relative to HST; see bottom right panels of Figure \ref{fig:2111_lens}, where several clumps -- both small in size and faint in flux -- are detected by eye in NIRCam observations, that were previously undetected in HST WFC3 imaging (in the case of SGAS2111, which seems to have many quenched regions, the ability to be able to pick filters sampling SEDs redwards of the Balmer Break is an additional benefit). This characterization of clumps directly impacts our ability to address both clump formation as well as overall mass assembly in SGAS2111.  

In the subsection below, we share our approach towards selecting clumps and modeling their properties, by using NIRCam observations as an example.

\subsubsection{Modeling the NIRCam PSF}

There are a number of effects that make measuring the NIRCam PSFs more challenging than the HST PSFs.  The most prominent of these effects are that the width and shape of the NIRCam PSF can vary spatially across the detector quite significantly \citep{nardiello22,berman24} and that the NIRCam wide-band filters used in this study are indeed quite wide, so changes in the diffraction limit as a function of wavelength introduce a measurable spectral dependence to the PSF. The spatial dependence disfavors empirical PSF measurements because it means that stars used to measure the PSF must be located near the arc.  It is unlikely that an appropriate number of reasonably bright, isolated stars will appear so close together in every field that we target.  The spectral dependence also disfavors empirical measurements because the stars that are averaged together to build an empirical PSF will likely have different spectral types, none of which accurately represent the observed spectrum of a clump at z$\sim$3 because they are all observed at $z=0$.

Models of the NIRCam PSF created using the commonly-used code, \texttt{STPSF} \citep{stpsf} (WebbPSF at the time of observations), are known to be too narrow to represent the PSF in level 3 combined NIRCam images and the documentation recommends using those models only in level 2 add \texttt{rate} or \texttt{cal} files. \citet{jipsf} attempted to solve this problem by creating PSF models with STPSF , inserting them into the cal files, and then processing them to final level 3 \texttt{i2d} products and measuring PSFs from those images.  This was generally successful for their purposes, but in testing this method with our data, we determined that further improvements were possible.

We created a suite of STPSF models, varying the subpixel positions of the centers and testing a variety of illuminating source spectra. We scaled these model PSFs to best match real stars in the Level 2 rate or cal files. We find that these effects together can introduce uncertainties of up to a few percent in the recovered flux and produce noticeable residual patterning after subtraction. The dominant source of uncertainty arises from the subpixel position, while the assumed source spectrum contributes at roughly the 1\% level. Furthermore, while \citet{jipsf} pointed out that the level 3 PSFs in real data are wider than those in level 2 images, we found that STPSF models of the NIRCam PSFs were also a little too narrow compared to real point sources in the level 2 images.  We found that this bias occurred across filters, detector locations, and datasets taken at different times.  We experimented with smoothing the STPSF model using a Gaussian kernel and found that smoothing by a roughly 0.5 pixel wide Gaussian consistently produces better matches to the real data (to be exact, we achieved the best fits with a gaussian with $\sigma=0.4717$).  This was remarkably consistent across filters and the time of the observation, so we interpret this as potentially an unmodeled detector-level effect.  It is possible that a smoothing kernel with a different functional form or some other type of transformation may further improve the fit and work on that is ongoing.

With all of this in mind, we create our PSFs for this work as follows.  We find the RA and Dec of the precise center of a pixel inside the arc in the level 3 images (often this pixel was chosen to be on or near a bright clump because details of the PSF, including the effects of spatial variation across the detector, affect photometry most the smaller and brighter the object is).  Then we create STPSF models at the precise subpixel location of that RA and Dec in each cal file and add them to a copy of that cal file (which first had each pixel set to 0), to the precision possible with the current NIRCam distortion models (typically about 0.1 pixel or better).  These are then smoothed with a Gaussian kernel with $\sigma=0.4717$ and run through stage 3 of the NIRCam pipeline to create a level 3 image of the PSF, centered at the center of the pixel we had chosen.  We then extract a region centered at that pixel and use it as our PSF for the work in this paper.  STPSF normalizes the PSF at infinity and this normalization is preserved.  We chose an illuminating source spectrum of an A0V star from \citet{ck04}, redshifted to match the redshift of the arc.

\subsubsection{Measuring Sizes}

As a demonstration, we measure the intrinsic effective radius of clumps for galaxies in our sample by fitting regions in the bluest JWST filters. Clumps are first selected by visual inspection through color images composed of several of the available NIRCam filters. This approach creates a lower limit for the number of total clumps in both objects as there is strong bias towards selection of only the brightest or most visually distinct clumps.

As an example, from our visual selection, we find 53 clumps in SGAS1110 and 49 clumps in SGAS2111 (see Figure \ref{fig:mag_size_pdf}). To estimate the intrinsic sizes of the clumps, we select NIRCam filter F115W as a reference for observed clump size due to its availability in both targets and low PSF FWHM (except for the few cases where data is only well-modeled in F150w, or for SGAS1110 alone, sometimes F182M). We create an image cut-out for each clump ranging in size from 5 x 5 pixels to 13 x 13 pixels depending on the observed clump size and proximity to other bright sources. Each cut-out is then individually fit with clumps modeled by a 2D elliptical Gaussian convolved with the instrumental PSF (see previous section) and the background modeled by a 2D polynomial function to the first degree; this is inspired from the analysis described in section 3.2 of \citealt{Messa2022}. In our free parameters, we include both the minor and major standard deviation axis to allow for an unbiased fitting to clump shape. For image cut-outs that include clumps too near to one another to exclude via a smaller cropping or with background modeling, we instead fit two Gaussians to the cut-out.

To estimate the effective radius of each clump, we first compute the circularized radius from the PSF-convolved Gaussian fit,

\begin{equation}
    R_{\rm cir} = \sqrt{\sigma_x \cdot \sigma_y},
\end{equation}

\noindent where $\sigma_x$ and $\sigma_y$ are the standard deviations along the minor and major axes of the fitted Gaussian, respectively. We then convert to a PSF-deconvolved observed radius by applying the half-width at half-maximum (HWHM) relation,

\begin{equation}
    R_{\rm obs} = R_{\rm cir}\sqrt{2\ln 2},
\end{equation}

\noindent and finally obtain the intrinsic (source-plane) effective radius by correcting for gravitational lensing magnification,

\begin{equation}
    R_{\rm eff} = \frac{R_{\rm obs}}{\sqrt{\mu_{\rm median}}},
\end{equation}

\noindent where $\mu_{\rm median}$ is the median magnification over the pixels composing the Gaussian profile of the clump.

See Figure \ref{fig:mag_size_pdf} for a distribution of magnifications and intrinsic (source plane) sizes for 102 clumps identified in SGAS2111 and SGAS1110. In these representative galaxies, we probe 10-200 pc regions across each primary image of the galaxies.

\subsubsection{Image plane fluxes via aperture photometry} \label{sec:aperturePhotometry}

Several pre-processing steps were carried out before measuring the photometry of the selected clumps.  SGAS2111-0114, is---at least relative to the rest of the LEGGOS sample---a rich cluster.  Simple aperture photometry of the arc would be influenced by significant intracluster light (ICL) and the flux of nearby cluster member galaxies.  Level 3 mosaics for some filters of SGAS1110+6549 showed evidence of a gradient in the sky background.  To deal with both of these issues, we used GALFIT \citep{Peng2002,Peng2010} to fit foreground contamination (i.e., ICL, cluster galaxies, other foreground galaxies) and sky pedestals and gradients.

Photometry was conducted using images that were PSF-matched to the filter with the largest PSF (F480M for SGAS1110+6549, F444W for SGAS2111-0114).  Matching kernels were created using \texttt{pypher} \citep{pypher}.  Fluxes were extracted from within the elliptical apertures described in \autoref{subsec:selection} using PSF-matched images.  When only a fraction of a pixel fell within an aperture, only that fraction of the flux in that pixel was included.  In order to calculate uncertainties, accounting for the effects of convolving the images with the kernels described above, we did the following.  We created 100 realizations of the data in each filter where the pixel values were drawn from Gaussian distributions centered at the value of that pixel in the science extension of the level 3, non-PSF-matched data, with standard deviation given by the value at the corresponding position in the error extension.  We then convolved those 100 realizations of the image with the PSF-matching kernel, and summed the fluxes inside the ellipse as described above.  At the same time, we also calculated backgrounds in apertures of the same size and shape, randomly placed in the blank regions of the sky near the arc (within a 150 pixel box centered on the clump in question, an area large enough to include blank sky pixels outside of the arc itself, but small enough to capture local variations in the sky).  Backgrounds were subtracted from the measured fluxes.  For each clump, we then adopted the median background-subtracted flux from those 100 realizations as the clump flux and the $16^{th}$ and $84^{th}$ percentiles are adopted as the uncertainties. We add an additional 3\% uncertainty on the flux (in quadrature) to account for PSF and PSF-matching uncertainties. We correct any measured photometry for Galactic reddening using the Schlafly \& Finkbeiner 2011 dust map \citep{Schlafly2011} on the NASA/IPAC Infrared Science Archive\footnote{\href{https://irsa.ipac.caltech.edu/applications/DUST/}{irsa.ipac.caltech.edu/applications/DUST/}}.


\begin{figure}
    \centering
    \includegraphics[width=\linewidth]{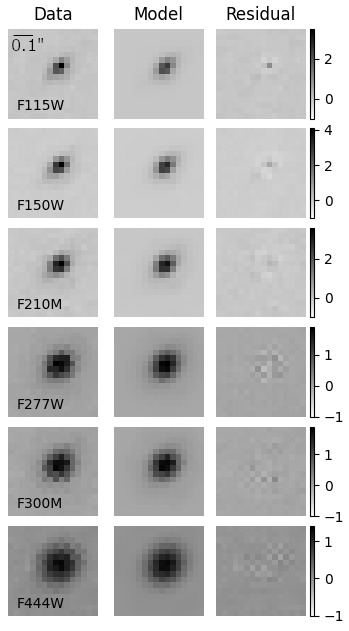}
    \caption{From left to right: The observed image plane NIRCam data for Clump 4 in SGAS2111, the forward-modeled image plane, and the residual (colorbar units in MJy sr$^{-1}$). At z=2.86, 0.1" corresponds to $\sim$750 pc, and a nominal magnification of 20$\times$ allows us to observe $\sim150$ pc regions.} 
    \label{fig:fomo}
\end{figure}

\subsubsection{Source plane measurements via forward modeling clump fluxes and morphologies}\label{sec:fomo}

Extraction of intrinsic properties like luminosity or size of small regions like clumps in distant galaxies is challenging.  The flux from those clumps is embedded within diffuse light of the host galaxy, potentially complicated further by structures like spiral arms or bars, or even more complicated features if the host galaxy is, for example, in the midst of an ongoing merger.  The telescope/instrument PSF further complicates the matter by blurring the light from those two components together---an especially important consideration for clumps, which are often similar in angular extent to (or even smaller than) the size of the PSF.  Detailed morphological decompositions using codes like GALFIT \citep{Peng2010} can separate these components using parametric models assigned to each of the physical components (i.e., each clump, diffuse light from the host galaxy, a central bulge, etc.) and solving for the parameters that produce models that when convolved with the PSF, best match the observed data.  In lensed galaxies, specifically, shear and differential magnification fundamentally change the task of morphological modeling in ways that demand more complicated approaches.

While parametric models like S\'{e}rsic profiles (and limiting cases like exponential, De Vaucouleur, and even Gaussian profiles) or King profiles may be well-suited to matching large scale galaxy light profiles or the clumps within them in unlensed galaxies, the distortions caused by lensing alter the observed light distributions and even introduce properties like curvature.  Lens models can enable source reconstructions, in which the effects of lensing are effectively reversed in order to show what the source galaxy would have looked like if it had not been lensed (i.e., in the ``source plane").  However, because the observed light profile of the galaxy is the ``true" profile convolved with the telescope/instrumental PSF, any transformation done to the data to reverse the lensing effect is also done to the PSF.  As a result, the ``unlensed" light profile of the galaxy in the source reconstruction is effectively convolved with a PSF that can vary significantly from location to location in the source plane (i.e., at different locations within the source galaxy).  There can, therefore, be quite large changes in the size, ellipticity, and even rotation of the effective PSF as a function of source plane position.

Though many authors have fit lensed galaxies in the image plane using parametric models like combinations of several to dozens of S\'{e}rsic profiles (e.g., \citealp{Lin2009, Khullar2021,Mowla2024,Cloonan2025}, among others) and accepted the uncertainties introduced by using parametric models that are known not to accurately mimic the underlying physical light profiles, a better approach is to use forward-modeling.  \cite{Johnson2017_1110paperI} used such an approach to model the HST data of one of the sources in this program, SGAS1110+6459. The basic idea is to make use of the fact that the actual, physical components of the galaxy are well-modeled by common parametric profiles in the undistorted (source) plane while the PSF is most stable and measurable (or modelable) in the image plane (where the actual light of the galaxy is distorted).  The process involves constructing a model of the source galaxy light profile using a combination of parametric profiles---Gaussians in the case of \cite{Johnson2017_1110paperI}---and using the lens models to ray trace those model profiles to the image plane.  There, the resulting lensed light profiles are convolved with the PSF (avoiding the need to account for extreme variations in PSF size, shape, and orientation that would have been necessary in the source plane) and compared to the observed data.  Models are then tweaked in the source plane until the difference between the model and observations has been minimized.

We follow a similar approach, though generalizing beyond what was done in \cite{Johnson2017_1110paperI}.  Rather than building the model of clumps in the source plane and a model of the more diffuse light of the host galaxy in the image plane, we model both components in the source plane.  We also allow for simultaneous fitting of the source plane galaxy model with a model of the background sky noise, in the image plane, that includes a pedestal as well as gradients in both the x- and y-directions.  Instead of Gaussian profiles in the source plane, we experimented with the more generalized S\'{e}rsic profile for both clumps and diffuse components.  While this produced models that matched the observed data in the relatively bright centers of clumps, we found that light in the more extended wings of the S\'{e}rsic profile was not well-constrained. This occurs for a number of reasons.  First, S\'{e}rsic profiles, formally, extend to infinity (though the surface brightness, of course, drops off fast enough that the integrated flux is finite) while the actual clumps and clusters of stars that we are attempting to model have a finite size.  Second, when modeling clumps near a caustic, light that extends into the wings of the S\'{e}rsic profile can extend to the other side of the caustic and not even appear in the image plane in some lens configurations (like the merging pair of images of SGAS2111-0114), thus providing no constraints on that light in the real data.  Extended light that does not cross the caustic can still be thrown to wide distances that may not be included in the fitting region in the image plane and be unconstrained by the data, or can have such low surface brightness that it is poorly constrained by the data, but can sum to a significant amount of light across a large enough area.  We found, by fitting a small region within SGAS2111-0114, which we describe below, that the contribution of flux in the extended wings of a source plane S\'{e}rsic profile can add a 15-20\% bias to the inferred fluxes relative to, for example, an image plane GALFIT decomposition.

To rectify this, we turned our attention to King profiles, which were originally developed to model star clusters---including globular clusters---in the Milky Way \citep{King1962,King1966}, suggesting that they may do a better job than S\'{e}rsic profiles of approximating the real physical objects (like clumps, some of the most highly magnified of which may even be globular clusters).  An additional benefit of this choice is that King profiles have an explicit truncation radius, eliminating the possibility of poorly-constrained extended wings creating the sorts of biases we saw when using S\'{e}rsic profiles.  We tested various combinations of King and S\'{e}rsic profiles, as well as either including or not including an overall pedestal and gradient using a highly magnified ($\mu \sim 30$) clump in SGAS2111 using this approach.  Ultimately a King profile was the best recovered clump flux.   We elaborate on this in \S~\ref{sec:fomoResults}.

For the purposes of this example, we have chosen to model a small region around a clump that lies close to the critical curve.  This clump was chosen because it is highly magnified, appears at least marginally resolved (has visible elongation in the bluest NIRCam bands), and is somewhat isolated which simplifies the required modeling. However, it does still have clearly visible diffuse light from the host galaxy around it, particularly in the longest wavelength bands, so it is not an oversimplification of the more general forward-modeling task for this program.

First, we masked the arc in the image plane and used GALFIT to model the sky, intracluster light from the lens, and several cluster galaxies and other foreground/background sources close to the arc that could have affected the eventual forward-modeled photometry.  We then selected a small region around the clump and created a source reconstruction on a pixel grid 100 times finer in both $x$ and $y$.  The size of the region was chosen to balance the computational inefficiency of modeling a large source plane pixel grid (keeping in mind the 100$\times$ finer pixel sizes which lead to a grid with roughly 10$^{4}$ times as many pixels) against the need to have off-clump sky pixels to constrain the lensed model light profiles.  For other targets, we may choose different ratios of image plane to source plane pixel size based on the magnification in order to balance computational cost against fine sampling of the source plane.

\begin{figure}
    \centering
    \includegraphics[width=\linewidth]{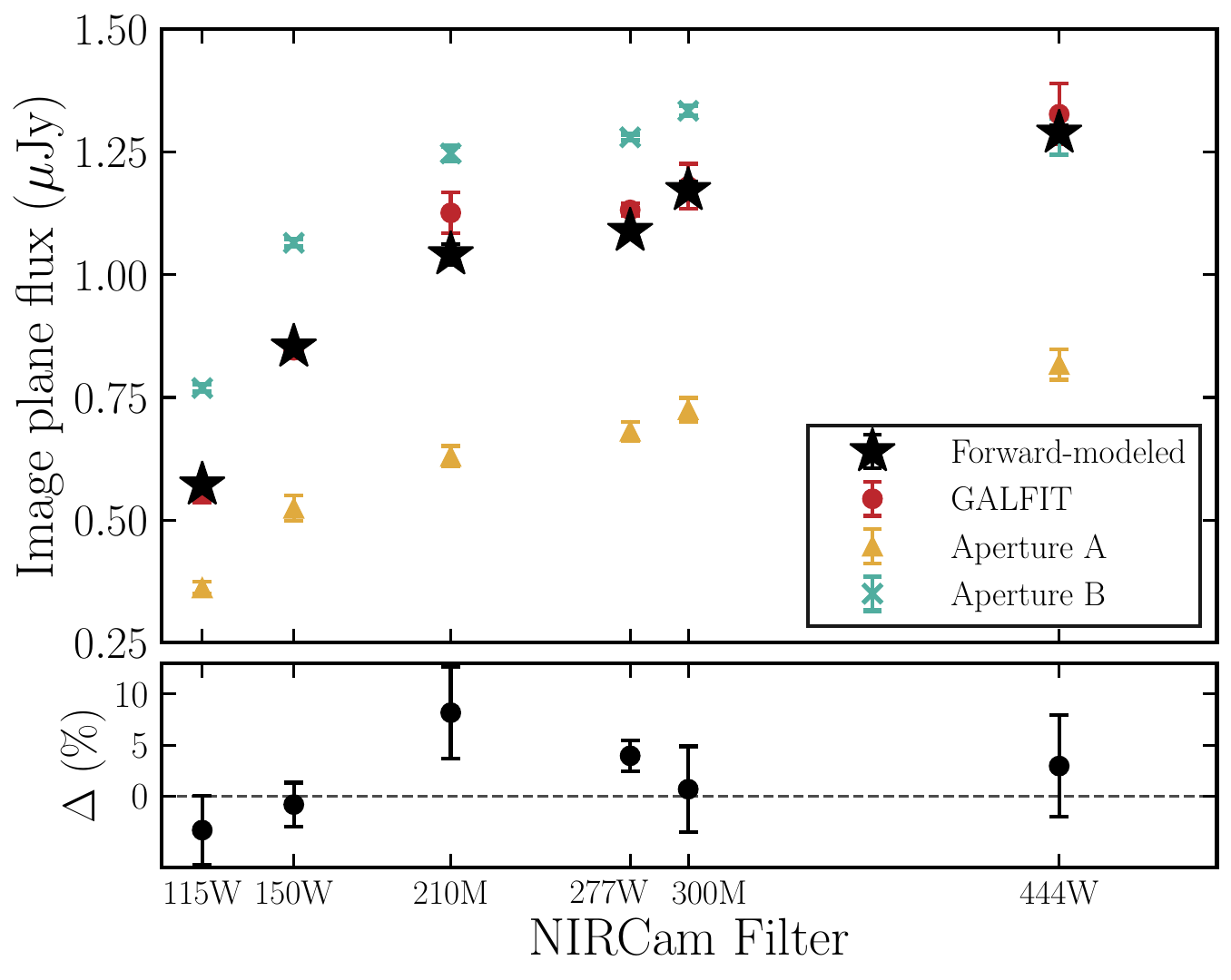}
    \caption{Comparison of photometry for Clump 4 in SGAS2111 derived from various aperture and model photometry methods The bottom panel shows the percentage difference between GALFIT-based model photometry vs. forward modeled measurements; the fiducial methodology matches the GALFIT performance.}
    \label{fig:fm_vs_galfit}
\end{figure}

In that source plane reconstruction, we approximated the centroid of the clump by eye.  We placed a King profile at that position.  For the diffuse light of the host galaxy, we experimented with adding a S\'{e}rsic profile in the source plane\footnote{We also note, for readers attempting to recreate this process, common Python packages for generating 2D S\'{e}rsic profiles (like those built into \texttt{astropy}) tend to do an inadequate job of normalizing the flux, particularly in the central pixel, for small profiles and for those with high S\'{e}rsic indices.  This can result in incorrectly normalized fluxes that are off by up to several orders of magnitude.  The structures one may wish to model in strongly lensed galaxies can be small enough to push against and beyond those limits, so one may wish to use a different implementation of the S\'{e}rsic profile.  While developing our pipeline, we used GALFIT to render source plane S\'{e}rsic profiles to avoid this issue.}.  We also tried a model with no source plane S\'{e}rsic profile, but including a pedestal and gradient in the image plane to do a local fit of the diffuse component.  We also tested a model containing both of a source plane S\'{e}rsic component and an image plane pedestal and gradient.  We allowed all parameters describing all model components to vary, though they were bounded.  We first solve for a best-fit model using \texttt{scipy.optimize.minimize}, to minimize the difference between the data and lensed, PSF-convolved model (weighted by the pixel-by-pixel uncertainties).  We then sample around that solution using an MCMC process, implemented using the Python package \texttt{emcee} \citep{Foreman-Mackey2013}.  We choose to use a deterministic minimizer first, in part because solutions are extremely sensitive to $x$ and $y$ position, especially when the clump is close to the critical curve and, in our experience, methods like gradient-descent find good positions faster than simply running an MCMC chain from the start.  With this optimized model in hand, we use MCMC to further explore parameter space, sampling around the best-fit parameters to find potentially better solutions and estimate uncertainties. A comparison of the photometry measured using this method to photometry measured using GALFIT models and two different aperture photometry approaches is shown in Figure \ref{fig:fm_vs_galfit} and further discussed in the forward modeling results section, \S~\ref{sec:fomoResults}.

\subsubsection{Using UMAP and HDBSCAN to cluster SEDs together}

One of the primary motivations of forward modeling of image plane/source plane ``idealized" clumps is to robustly differentiate the clump fluxes and morphologies from the surrounding diffuse light. In \cite{Ross2026}, we conduct clump identification by grouping pixels by spectra/spectral energy distributions. This study employs Uniform Manifold Approximation and Projection (UMAP) to conduct IFS spaxel-based dimensionality reduction, followed by HDBSCAN-based unsupervised clustering, leading to distinct groups of pixels -- some of them part of bright clumps -- in \eleventen.

\subsection{Stellar population synthesis modeling -- photometry vs spectrophotometry} \label{sec:sps}

We conduct stellar population synthesis (SPS) modeling via Bayesian SED fitting and infer individual star formation histories to identify recent epochs of SF bursts (and quenching). We model the full NIRCam plus NIRSpec spectrophotometric data for all 8 spectroscopic targets. The details of our modeling methodology are as follows.

For this analysis, we utilize flexible star formation history modeling with the SED fitting framework \code{prospector} \citep{Johnson2017, Leja2017,Johnson2021}. \code{Prospector} uses the Flexible Stellar Population Synthesis (FSPS) stellar population synthesis models \citep{Conroy2009, Conroy2010}, the MILES spectral library \citep{Sanchez-Blazquez2006,Falcon-Barroso2011}, and the MIST isochrones \citep{Choi2016, Dotter2016}. We use non-parametric SFHs and implement a flexible age bin model, which utilizes fixed time/age bins at early times, flexible bins that each form the same amount of total stellar mass at intermediate times, and a final age bin with a flexible age boundary (closest to the epoch of observation); this model is optimized to constrain mass assembly in recently quenched galaxies (see \citealt{Suess2022b,Setton2023}). This scheme is designed to recover both bursty SF and quenching timescales with spectro-photometric data \citep{Suess2022b, Suess2022a}. 

In our fiducial model, we define the two fixed time bins at early times, such that they sample 40\% of the age of the Universe (0.4 $\times$ t$_{univ}$), while simultaneously fitting for the redshift, $z_{spec}$. The youngest age bin is sampled in the analysis in the range [0,0.3 $\times$ (t$_{univ}$)], as well as the flexible age bins. Moreover, we fit for redshift by sampling the range [$z_{spec}$ - 0.05, $z_{spec}$ + 0.05], to allow for an extended posterior distribution for the redshift, including accounting for Doppler broadening and wavelength calibration uncertainties. 

We assume a \cite{Chabrier2003} initial mass function. We adopt the \cite{Kriek2013} dust law with $A_v$ and dust index as free parameters, with doubled attenuation around young ($<10^7$ yr old) stars \citep{Wild2020,Suess2022a,Setton2023}. We fix the shape of the IR SED following the \cite{Draine2007} dust emission templates, with $U_\mathrm{min}$ = 1.0, $\gamma_e$ = 0.01, and $q_\mathrm{PAH}$ = 2.0., and marginalize over these parameters. We also marginalize over nebular line and continuum emission in our fiducial models (while testing our results against \textsc{Cloudy} based nebular emission line estimates). Finally, we also fit for both stellar and gas-phase metallicity (independent of each other). This analysis does not include any priors derived from either a mass-metallicity relation (MZR) or SFH priors from semi-analytical galaxy models (such as the ones encoded in UniverseMachine; \citealt{Behroozi2019}), as the SED fitting is conducted in the image plane of each galaxy, without any strong lensing magnification correction (which is implemented post-SED fitting). We use the \texttt{dynesty} dynamic nested sampling package \citep{Speagle2020} to sample the posterior distributions.   

We estimate de-lensed SFRs and stellar masses for the sample using the posterior distributions of the non-parametric star formation history, the total stellar mass formed (corrected for mass loss due to stellar winds, supernovae, etc.) and the multiplicative lensing magnification.

We also fit the same datasets with multiple flavors of SED models with varying definitions of age bins and varying ranges of metallicity and dust priors, primarily to rule out the impact of age bin definition and the age-metallicity-dust degeneracy on the duration of SF/quenching epochs, burst times and current SFRs in our clumps.

\begin{deluxetable*}{lccccc}[!ht]
\tablecaption{\label{table:photometry}Comparison of Photometry Methods}
\tablecolumns{6}
\tablehead{
\colhead{Filter} & \colhead{Flux ($\mu Jy$)} & \colhead{Flux ($\mu Jy$)} & \colhead{Flux ($\mu Jy$)} & \colhead{Flux ($\mu Jy$)} & \colhead{Source Plane Flux ($\mu Jy$)} \\ 
~\vspace{-7mm}\\ 
 & \colhead{(Forward-modeled)} & \colhead{(GALFIT)} & \colhead{(Aperture-A)} & \colhead{(Aperture-B)} & \colhead{(Forward-modeled)}\\
\colhead{(1)} & \colhead{(2)} & \colhead{(3)} & \colhead{(4)} &\colhead{(5)} & \colhead{(6)}}
\startdata
F115W & $0.572\pm0.007$ & $0.553\pm0.018$ & $0.362\pm0.012$ & $0.769\pm0.007$ & $0.0202\pm0.0016$\\
F150W & $0.853\pm0.013$ & $0.846\pm0.013$ & $0.524\pm0.026$ & $1.065\pm0.007$ & $0.0299\pm0.0023$\\
F210M & $1.041\pm0.021$ & $1.126\pm0.041$ & $0.630\pm0.021$ & $1.247\pm0.015$ & $0.0364\pm0.0030$\\
F277W & $1.089\pm0.010$ & $1.132\pm0.013$ & $0.681\pm0.019$ & $1.280\pm0.006$ & $0.0383\pm0.0027$\\
F300M & $1.172\pm0.017$ & $1.180\pm0.046$ & $0.725\pm0.024$ & $1.334\pm0.010$ & $0.0409\pm0.0029$\\
F444W & $1.289\pm0.015$ & $1.327\pm0.062$ & $0.817\pm0.031$ & $1.281\pm0.037$ & $0.0454\pm0.0041$\\
\enddata 
\tablecomments{Fluxes for Clump 4 in SGAS2111 using different methods of photometry. (1) NIRCam filter. (2) Image plane flux using the forward-modeling process described in \S~\ref{sec:fomo}. (3) Image plane fluxes using GALFIT models.  (4) Image plane fluxes using the aperture photometry method described in \S~\ref{sec:aperturePhotometry}. (5) Image plane fluxes using the aperture photometry method described in \citet{Mowla2022} (6) The source plane flux based on the forward-modeling method, with magnification uncertainties included. All fluxes are in reported in $\mu Jy$. Image plane fluxes are not corrected for Milky Way reddening, but the reported source plane flux in column 6 is and can be taken as the intrinsic flux of this clump.}
\end{deluxetable*}

\subsection{Nebular Line Emission Modeling}

Our approaches are an amalgam of \cite{welch2025} and \cite{hutchison2026} (inspired by ERS TEMPLATES; \citealt{Rigby2025}), as well as the analyses in \cite{riverathorsen2025arxiv}. We summarize those methodologies here.

To map the emission lines, we first model the stellar continuum using the \texttt{continuum} tools provided by the TEMPLATES ERS team\footnote{\href{https://github.com/JWST-Templates/jwst_templates}{github.com/JWST-Templates/jwst\_templates}} and applied as follows: first masking out wavelength slices within 1000~\kms\ of each detected emission line, interpolating over the masked areas, and smoothing the continuum using a boxcar convolution with a boxcar wavelength window of 100\ang\ in the rest-frame. We subtract the resulting continuum fits for each spaxel to derive a continuum-subtracted cube.

We then fit the emission feature on a spaxel-by-spaxel basis using the \texttt{curve\_fit} function in the \texttt{scipy} python package \citep{Virtanen.2020.SciPy}. We fit isolated emission lines with a single Gaussian. For line complexes, including the \oiii\ \lam\lam4960,5008 doublet or blended features such as \nii+\ha, we model the lines together as a doublet or triplet of Gaussians with the line widths and line centroids tied to one another. When fitting the \oiii\ doublet, we fix the \oiii\ \lam5008 / \lam4960 flux ratio at the theoretical value of 2.98 \citep[e.g.,][]{storey2000,Osterbrock.2006}.  When fitting the \nii+\ha\ complex, we fix the flux ratio of \nii\ \lam6585 / \lam6550  at the theoretical value of 2.8 \citep[e.g.,][]{Osterbrock.2006}.

For line doublets such as \sii, where the line ratios can vary depending upon gas density and temperature \citep{Berg.2018,Osterbrock.2006}, we allow the flux ratio to vary within the theoretical range (approximately 0.4--1.4 for \sii\ and 0.3--1.5 for \oii; e.g., \citealt{Berg.2018,Sanders.2016}). We fit unresolved \oii\ with a single Gaussian in the spatially resolved maps, and with two Gaussians to the spatially-integrated (or ``global'') \oii\ line profile. We correct the NIRSpec/IFS data for Galactic reddening using the Schlafly \& Finkbeiner 2011 dust map on the NASA/IPAC Infrared Science Archive\footnote {\href{https://irsa.ipac.caltech.edu/applications/DUST/}{irsa.ipac.caltech.edu/applications/DUST/}}.

\begin{figure*}
    \centering
    \includegraphics[width=0.85\textwidth]{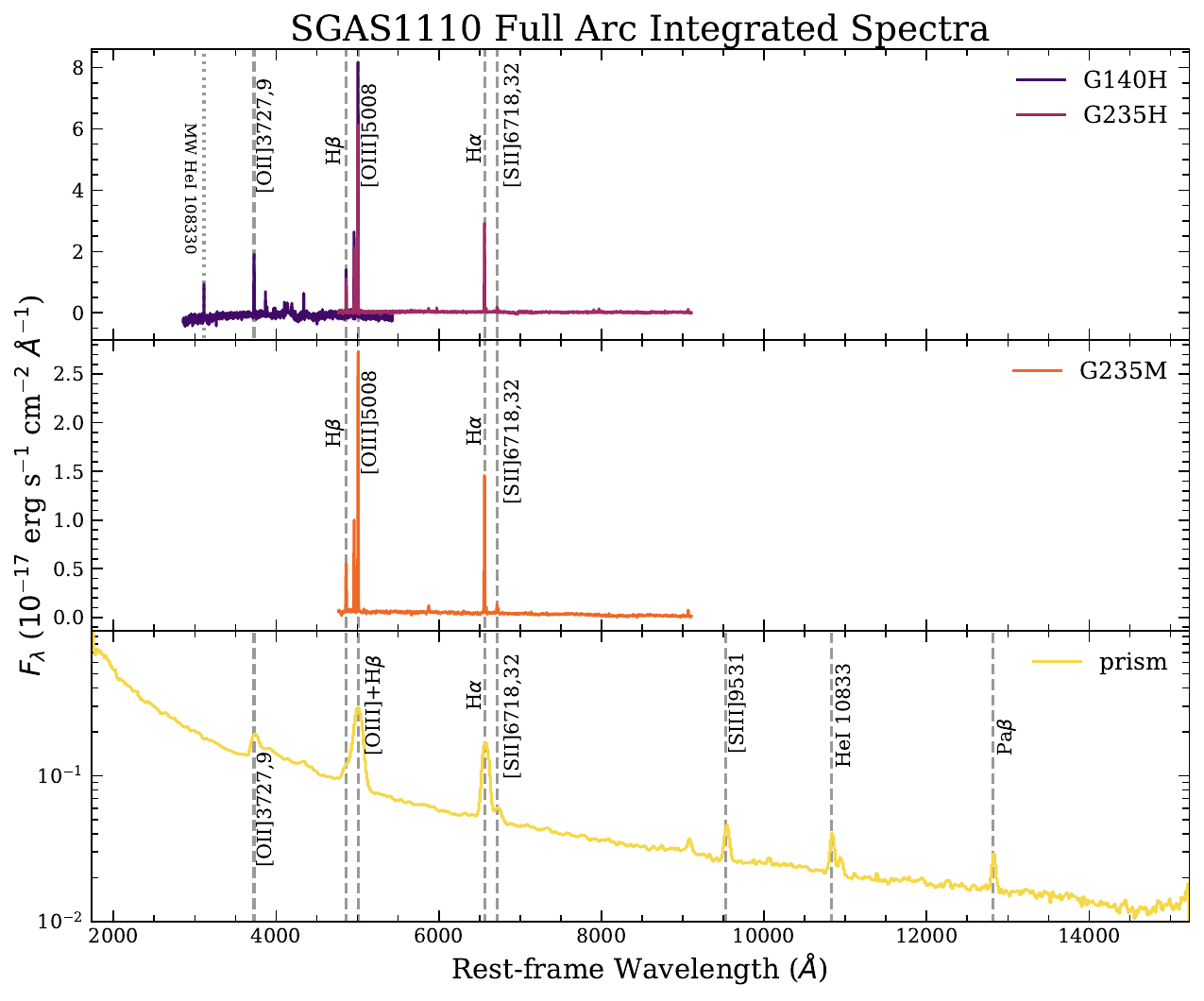}
    \caption{Integrated spectra from NIRSpec/IFS across the central (primary) image of the lensed arc; 
    this is one of the highest redshift legacy dataset comprising observations from all NIRSpec modes -- high-resolution H grating (top), medium-resolution M grating (middle), and low-resolution prism with the widest wavelength coverage (bottom).} 
    \label{fig:sgas1110_allspec}
\end{figure*}

\begin{figure*}
\centering
    \includegraphics[width=0.8\textwidth]{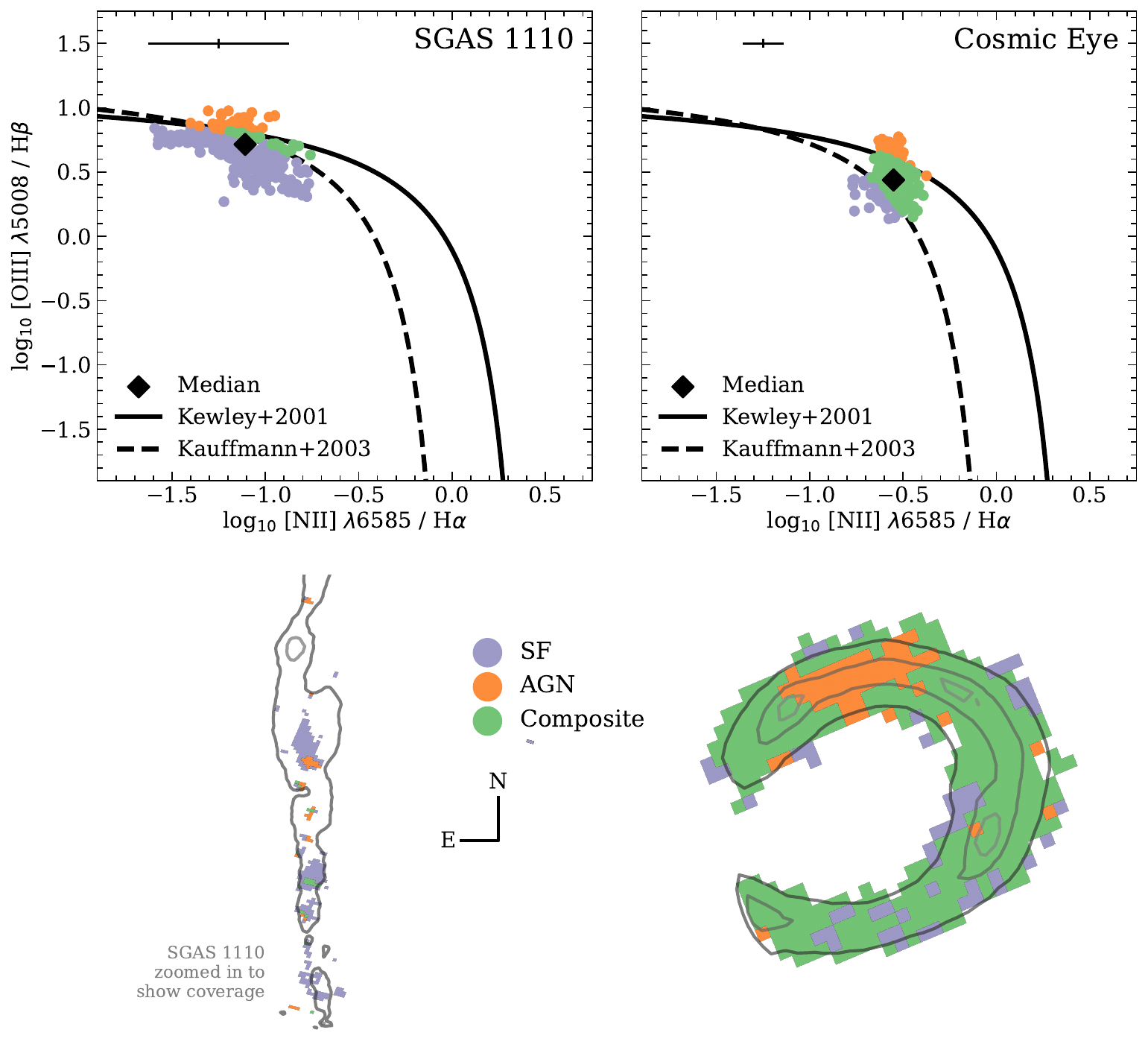}
    \caption{The \nii-BPT diagrams for the \eleventen\ (left) and \cosmiceye\ (right). The top panels show the pixel-by-pixel \nii-BPT diagrams, with both galaxies exhibiting line ratios consistent with star forming (purple), composite (green), and AGN (orange) regions as defined by the \cite{Kewley2001} and \cite{Kauffmann2003} diagnostics. The black diamond shows the median line ratios of all pixels in the galaxy, and error bars in the top left of each panel show the median uncertainties on each line ratio. The bottom panels show the image plane classifications with \ha\ flux contours overlaid (gray). }
\label{fig:bpt_gallery}
\end{figure*}

\begin{figure*}
\centering
    \includegraphics[width=0.89\textwidth]{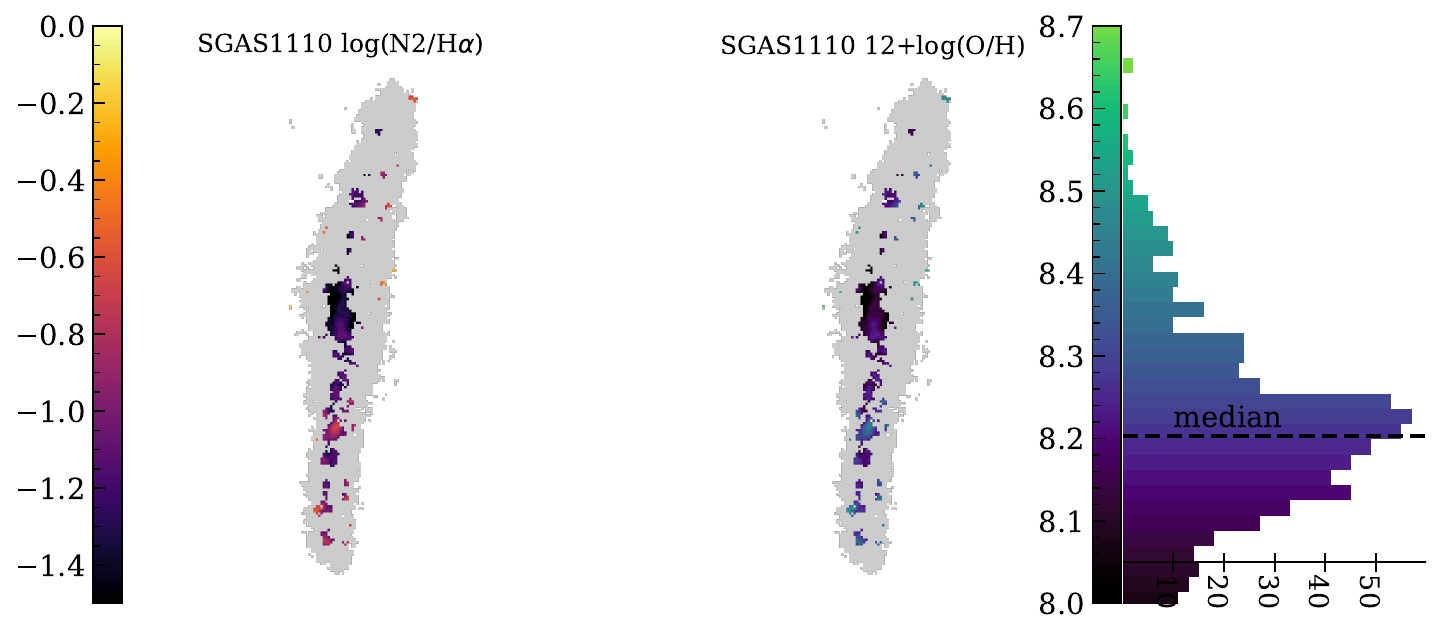}
    \includegraphics[width=0.89\textwidth]{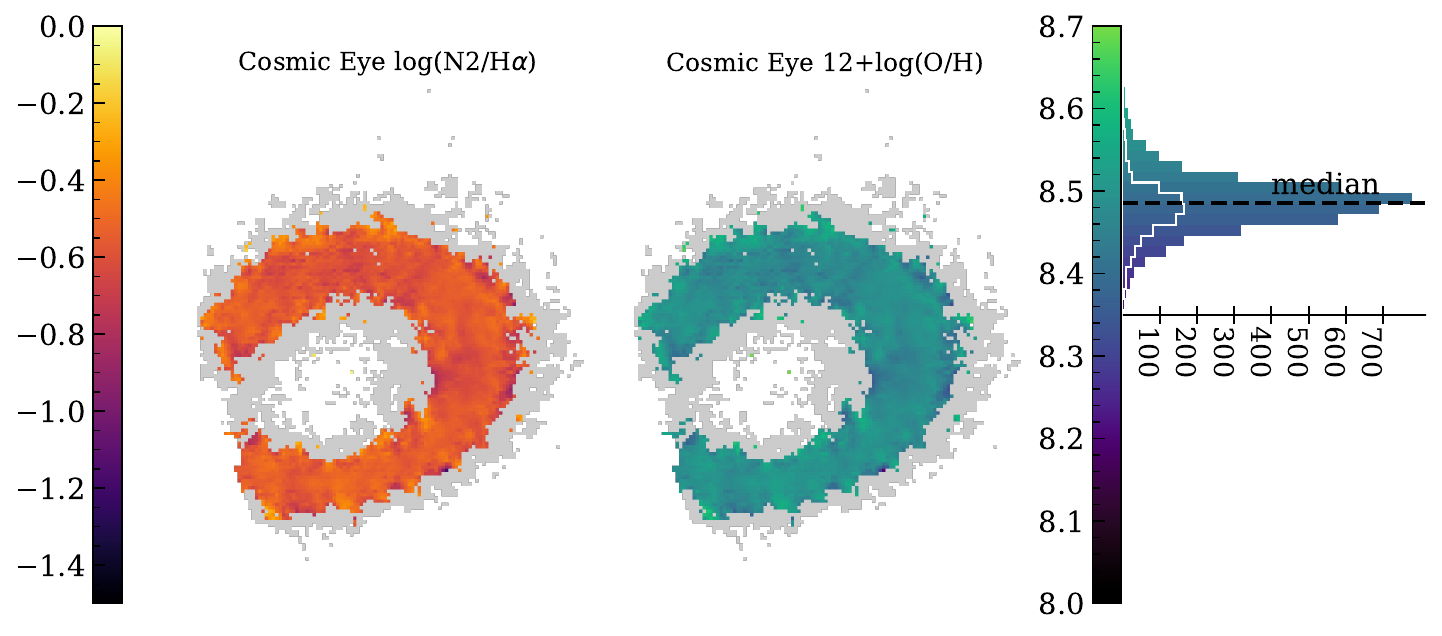}
    \includegraphics[width=0.89\textwidth]{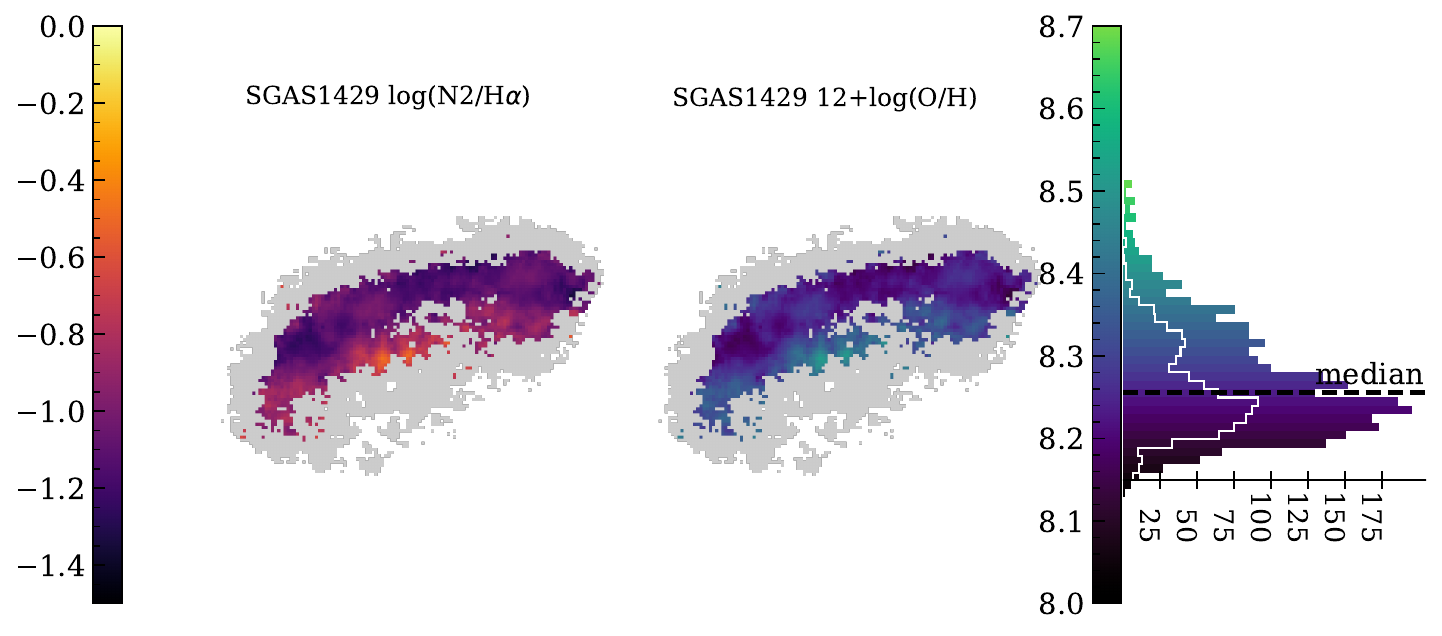}
    \caption{Maps of the N2 line ratio (left column) and N2-derived metallicity (right panel) are shown for three LEGGOS targets: \eleventen\ (top row), the Cosmic Eye (middle row), and SGAS1429 (bottom row). Histograms to the right of each metallicity map show the distribution of metallicity for individual spaxels in each source, with the median value indicated with a dashed black line. The white histograms plotted for the Cosmic Eye and SGAS1429 show the distribution of metallicities for spaxels within a single unique image of the multiply-imaged source. \eleventen\ does not include this single-image histogram because the full map covers only one image of the source. We see clear variations in metallicity across the source for \eleventen\ and SGAS1429, while the Cosmic Eye shows minimal variation in metallicity from N2. }
    \label{fig:n2ha_gallery}
\end{figure*}

\section{Initial Results and Discussion} \label{sec:discussion}

\subsection{Forward-modeling of NIRCam data for SGAS2111} \label{sec:fomoResults}
For clump 4 in SGAS2111, modeled as described in \S~\ref{sec:fomo}, the best residuals were found when including a pedestal and gradient in the image plane and only a King profile in the source plane.  In this case, because the sky had already been estimated and flattened in the GALFIT pre-processing step, the pedestal and gradient managed to approximate the local diffuse light of the host galaxy.  Adding a S\'{e}rsic profile in the source plane did not noticeably improve the residual, despite adding seven more parameters to fit. The results we report are, therefore, based on the model made from a single King profile, plus an image plane pedestal and $x$ and $y$ gradients. \autoref{fig:fomo} shows the data, model, and residual for this clump in the six NIRCam filters.  Drawing randomly from the MCMC chain and calculating the image plane flux of the clump, we find that the image plane fluxes generally agree well with fluxes inferred from GALFIT models of this arc in the image plane. However, they are systematically higher than those inferred from the aperture photometry method described earlier, by about 35\%, an offset that appears fairly consistent with wavelength. Comparing to another aperture photometry methods (see \autoref{table:photometry}), the one used for clumps in the Sparkler in \citep{Mowla2022}, we find a similarly-sized offset in the \textit{opposite} direction (except, notably for the F444W image) where forward-modeled fluxes are systematically lower than this aperture photometry. This may not be especially surprising, as those authors also acknowledge the difficulty in, for example, properly correcting for other light from the host galaxy or foreground cluster when performing aperture photometry. Ultimately, any aperture photometry method will have uncertainties enter due to measurement or modeling of the PSF, estimation of the matching kernel, estimation of contamination from the background (including the sky, diffuse light from the host galaxy, intracluster light, etc.) flux, and the difficulty of determining an aperture correction for a non-point-source. Forward-modeling removes the need to create matching kernels and simultaneously models the detailed light profiles of both the clumps and any contaminating light from the rest of the host galaxy, the sky, or foreground or background sources, minimizing the contributions of several of these sources of uncertainty, much in the same way that GALFIT modeling does. The agreement between GALIFT fluxes and forward-modeled fluxes is thus an indication that our forward-modeling process is accurately extracting the clump fluxes. Though image plane morphological decompositions, like GALFIT, do not directly yield information about source plane sizes, axis ratios, and positions, as these are not preserved in the image plane, forward-modeling does. An initial test suggests that sizes can be constrained to about 10\% or better, for example, comparable to or better than typical magnification uncertainties, but a full analysis of clump sizes and the systematics involved in determining them is left to a future paper. Forward-modeling also produces full posterior distributions for the values of these clump characteristics, something that is extremely difficult to do with any method based in the image plane.

Demagnified fluxes and uncertainties were estimated as follows.  215 realizations of the lens model were drawn from the MCMC chain produced by \texttt{lenstool} \citep{Jullo2007}, representing the range of possible lens models (and hence magnifications) that sample the parameter space around the best-fit model.  For each filter, we drew 1000 forward-modeled clump profiles from the MCMC chain using the best-fit lens model.  The image plane model light profile of the clump only (i.e., the lensed King profile) was created for each of these 1000 models.  Each one was then paired with a randomly selected lens model from the set of 215 allowed models (with replacement).  The image plane fluxes were divided, pixel by pixel, by the magnification predicted by the randomly selected lens model.  Importantly, this was done without convolving with the PSF.  Convolving with the PSF would scatter light away from its real, physical location (or, rather, our best model of its physical location).  Because magnification varies as a function of position, this should be avoided.

In this case, the posterior distributions of fluxes and magnifications were both quite symmetric and well-described by Gaussian distributions, so we adopted the median (50$^{th}$ percentile) demagnified flux as the inferred flux, and estimate uncertainties using the standard deviation of the distribution. In this way, we are able to jointly estimate uncertainties due to both our ability to forward model the light profiles and our ability to model the lens.  For this particular clump, the combined uncertainty is about 9-10\%, dominated by the magnification uncertainty.  Assuming an additional 2-3\% uncertainty due to our PSF model, and adding it in quadrature, does not change this by more than about half a percent.  This method is also significantly cheaper computationally and much faster than running new MCMC chains with each of the randomly selected lens model realizations, partly because slight changes in the model can create larger changes in the predicted $x$ and $y$ positions, requiring a new initial optimization before starting each MCMC chain, which often requires human intervention and cannot be reliably automated.  While in this particular case, the posterior distributions of fluxes and magnifications were approximately Gaussian, this may not always be the case for all lensed clumps, so in future work it may be more appropriate to report best-fit fluxes and $16^{th}$ and $84^{th}$ percentile fluxes even though they are effectively equivalent in this example.

\subsection{Emission Line Properties}

The following subsections describe various physical properties of gas and stars within LEGGOS sources derived via emission line measurements. As previously noted, one advantage of IFS over single slit spectroscopy is the ability to both spatially-resolve individual regions in a single source while also comparing to the globally-averaged properties (obtained by summing all IFS pixels together into a single spectrum).  Figure \ref{fig:sgas1110_allspec} shows one example of how IFS allows the same level of spectroscopic detail as slitted spectroscopy, while the following sections explore the power of IFS spatially-resolved science.
Additionally, Figure \ref{fig:sgas1110_allspec}'s 1D spectroscopy 
is one of the highest redshift legacy dataset comprising observations from all NIRSpec grating modes -- high-resolution H grating (top), medium-resolution M grating (middle), and low-resolution prism with the widest wavelength coverage (bottom).

\subsubsection{Ionizing Source Diagnostics}

Ratios of strong emission lines in the rest-frame optical are commonly used tracers of ionizing sources \citep[e.g.,][]{Baldwin1981,Veilleux1987,Kewley2001,Kauffmann2003,Kewley2019,Cleri2025}. By leveraging line ratios that trace important physical properties such as the ionization parameter and the gas-phase metallicity, commonly used emission line ratio diagnostics including the \nii-``BPT'' diagram \citep{Baldwin1981} were created to distinguish star-forming regions and hosts of accreting supermassive black holes. 

The \nii-BPT diagram compares the $\oiii~\lambda5008/\hb$ ratio to $\nii6585/\ha$. These diagnostics were designed to satisfy the following criteria: (1) the line ratios were made up of strong, easily detectable lines, (2) lines that are significantly blended should be avoided, (3) wavelength separation should be small to limit effects of dust attenuation and instrumental calibrations, (4) ratios of a forbidden line to a Balmer line are preferred to limit abundance sensitivity for a given metal, and (5) lines used in these diagnostics should all be readily accessible with current instrumentation \citep[i.e., limited to the rest-frame optical;][]{Veilleux1987}.

Figure \ref{fig:bpt_gallery} shows the \nii-BPT diagram with the \cite{Kewley2001} and \cite{Kauffmann2003} diagnostics for each pixel of SGAS1110 and the Cosmic Eye in which the S/N ratio for all four emission lines is $>3$. We also show the image plane pixel-to-pixel classifications with contours showing the \ha-detected regions. For SGAS1110, the \nii-BPT classifications find 80\%/6\%/14\% star-forming/composite/AGN. For the Cosmic Eye, the \nii-BPT classifications find 14\%/73\%/13\% star-forming/composite/AGN. 

We interpret the line ratios of SGAS1110 as arising from high-ionization parameter ionized regions, similar to observations of galaxies at higher redshifts \citep[e.g.,][]{Ubler2023,Scholtz2025,Trump2023,Backhaus2024,Backhaus2025,Mazzolari2024,Clarke2026,Cleri2026} and theoretical predictions of such systems \citep[e.g.,][]{Cleri2025,Richardson2025}.

The spatially-resolved spectroscopy from LEGGOS NIRSpec IFU observations is able to capture the diversity of line-emitting regions on the sub-galactic scale in a way that remained inaccessible at high-redshifts prior to JWST. The ionizing source diagnostics of the full LEGGOS sample will be analyzed further in an upcoming work (Luettgenau et al., in preparation).

\begin{figure*}[htb!]
    \centering
    \includegraphics[width=\textwidth,trim=0 50 0 40, clip]{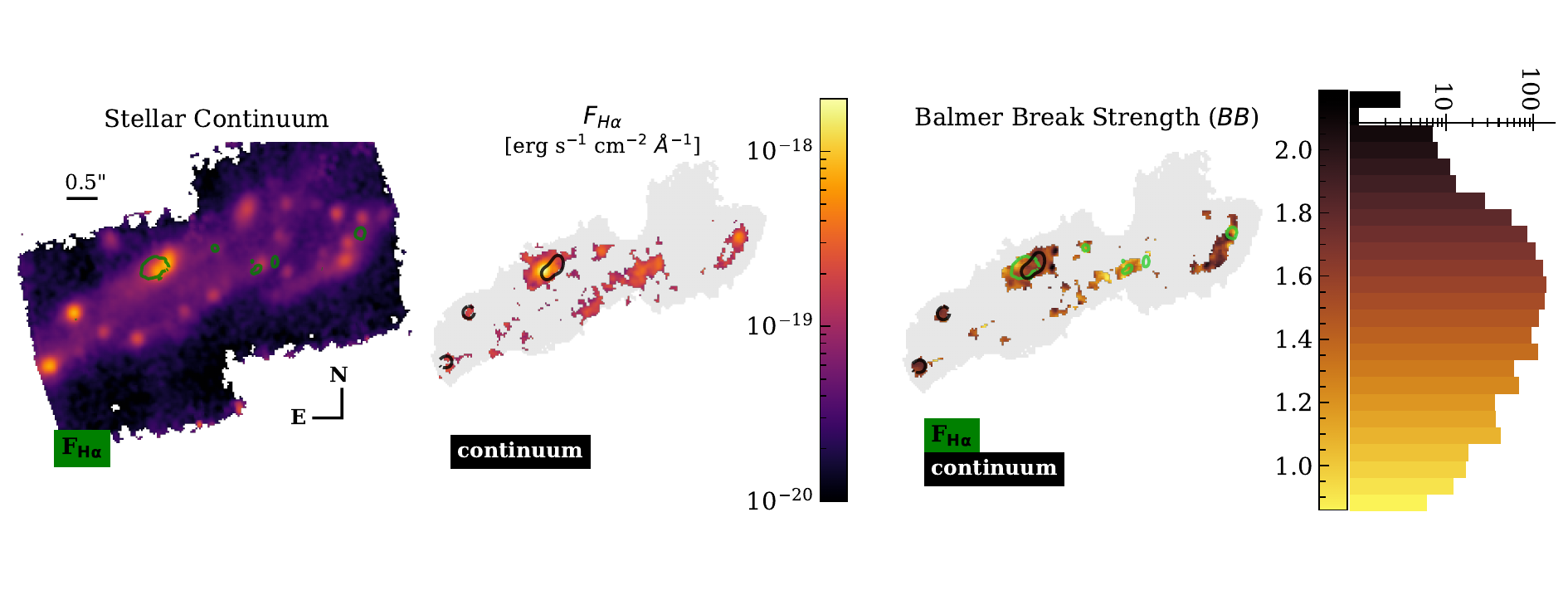}
    \caption{Selecting clumps/star clusters for investigation with LEGGOS NIRSpec/IFS prism observations. (\textit{left}) Stellar continuum image of SGAS 2111 from IFS pointings. The green contours represent regions of higher \ha\ emission ($F_{H\alpha} = 3\times10^{-19}$ erg s$^{-1}$ cm$^{-2}$ \ang$^{-1}$).  (\textit{center}) Flux map of the \ha\ emission line, with black contours representing regions of higher stellar continuum (with the same flux limit, $F = 3\times10^{-19}$ erg s$^{-1}$ cm$^{-2}$ \ang$^{-1}$). (\textit{right}) Map of the strength of the Balmer Break feature.  The black and green contours are the same as the previous panels. The histogram mapped to the colorbar represents the number and diversity of Balmer Break strengths across this galaxy.  These three panels represent key indicators that have the potential for differentiating unique physical regions within an individual galaxy.}
    \label{fig:2111_bb_ha}
\end{figure*}

\begin{figure*}
    \centering
    \includegraphics[width=0.95\textwidth]{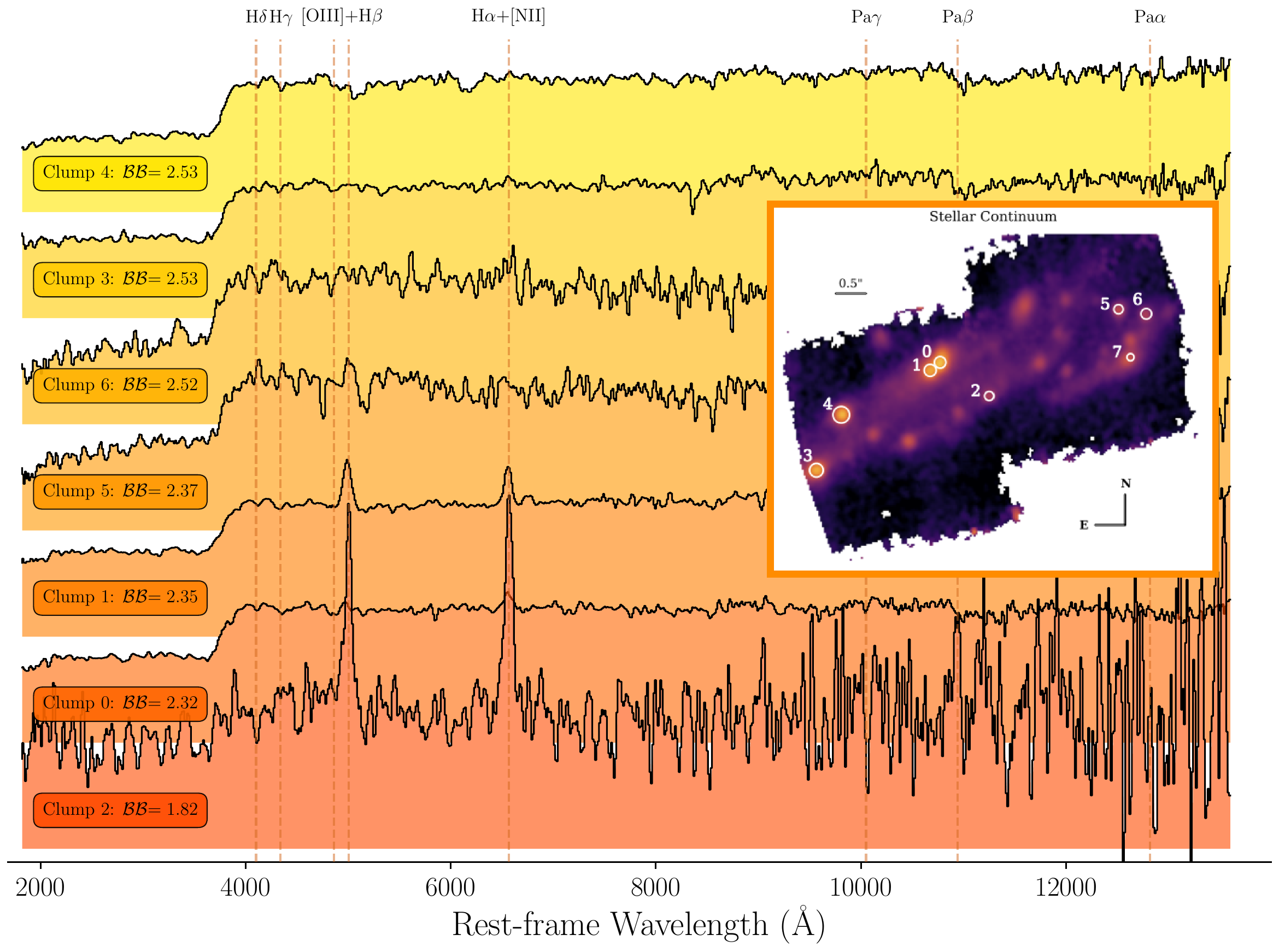}
    \caption{A gallery of spatially resolved JWST NIRSpec/prism spectra of visually identified "clumps" in SGAS2111 ($z=2.86$), plotted in decreasing order of Balmer break strength. Dotted vertical lines show prominent emission and absorption features from rest-frame $0.4-1.3\mu m$. Inset shows the locations of all seven spectra (CLumps 0-6) spatially on a map of stellar continuum emission from NIRSpec/IFS.} 
    \label{fig:joyplot_spectra_sgas2111}
\end{figure*}

\subsubsection{Gas-phase Metallicities}

We present spatially-resolved gas-phase metallicity maps for three targets - SGAS1110, the Cosmic Eye, and SGAS1429. 
We calculate the metallicity ($12+\log(\textrm{O/H})$) using the N2 ratio, defined as N2 $= \frac{[\textrm{NII}]\lambda6585}{\textrm{H}\alpha}$. 
To convert from the N2 ratio to metallicity, we use the empirical calibration of \cite{Marino2013} because it is based on direct-method metallicity measurements of individual HII regions, which are more analogous to the small spatial scales probed in our lensed galaxy sample, rather than on full galaxy spectra \citep[as in e.g.,][]{Curti2020,Sanders2025}. 
Additionally, the \cite{Marino2013} calibration does not rely on simultaneous constraints of the ionization parameter, as in the photoionization-model-based calibrations of \cite{Kewley2019}, thus simplifying this exploratory analysis. 
The N2 relation of \cite{Marino2013} is calibrated in the range $-1.6 < \log(\textrm{N}2) < -0.2$. 

In this initial discussion of LEGGOS results, we focus on a subset of the sample galaxies. All spaxels in SGAS1429 with well-measured N2 fall within the calibrated range of the \cite{Marino2013} relation. Only 3/4076 spaxels in the Cosmic Eye with well-measured N2 fall outside this range. For \eleventen, 18/679 spaxels have measured N2 below the lower limit of $\log(\textrm{N}2) = -1.6$. For these spaxels, we extrapolate the \cite{Marino2013} relation, accepting the significant systematic uncertainty that comes with extrapolation for this small fraction of spaxels. The measured N2 ratio and the derived metallicity are plotted in Figure \ref{fig:n2ha_gallery}. 

We see clear differences between the gas-phase metallicity conditions within and across these three targets. SGAS1110 is the lowest metallicity galaxy of the three highlighted here, and thus unsurprisingly shows the lowest detection rate of the \niib\ emission line. 
However, we see a broad distribution of metallicities inferred from this weak \niib\ in SGAS1110, with some clumps having higher metallicities than others. 
Additional study is ongoing to fully assess the variation in metallicity and other gas-phase physical conditions within this complex lensed arc (Lamprou et al., in prep.). 

The Cosmic Eye is at a higher metallicity, and thus has a much larger \niib\ detection rate. 
With this arc, we can successfully study the metallicity variation across the full extent of the galaxy. 
We find that the metallicity map is uniform spatially, with very little variation between different regions of the galaxy. 
Additionally, the distribution of metallicities appears to be nearly Gaussian, indicating that the variations we see are more likely due to statistical fluctuations rather than actual physical differences between regions. 
Such a flat metallicity profile has been observed in other galaxies at Cosmic Noon \cite[e.g.,][]{Curti2020klever}, and could be interpreted as evidence for stronger feedback inducing more efficient mixing of metal-enriched gas in these galaxies \citep[e.g.,][]{Ma2017metalgradient,Hemler2021TNGmetals}. 
However, we interpret these preliminary results with caution. 
Observations of the $z\sim 5$ Waz Arc found that the N2 metallicity had significantly less variation than other strong-line indicators \citep{hutchison2026}, while direct-method oxygen abundance measurements in the lower-redshift galaxy SGAS1723 found that all strong-line methods show less variation than direct-method metallicities (Olivier et al., in prep).
Additional analysis involving multiple metallicity tracers and additional gas-phase physical conditions is left for future work. 

SGAS1429 is at a lower metallicity than the Cosmic Eye, but has many more \niib\ detections than SGAS1110. 
Interestingly, this target shows some clear variations in metallicity based on the N2 ratio.
In the main arc, we see slight variations between the bright multiply-imaged clumps and the fainter diffuse structure of the galaxy. 
The clumps have higher N2 ratios, indicating higher metallicities than the surrounding diffuse components. 
We note that these clumps also appear redder than the blue diffuse arc, possibly indicating that these are older stellar populations which have had more time to enrich their surrounding gas. 
Additionally, we see that the secondary arc on SGAS1429 is at a higher metallicity than the primary arc. 
Again, this arc shows redder colors in the NIRCam images, possibly indicating an older stellar population which has enriched the surrounding gas. 
Interestingly, the N2 metallicity map shows two prominent clumps of high N2 in a region where no obvious structure exists in the NIRCam imaging. 
Additional investigation will be required to fully understand the structure of this galaxy, and is left for future work. 

\subsection{Quenching and Starbursts in LEGGOS Clumps} \label{sec:sps_results}

Using \twentyoneeleven\ as an exhibit for our star formation history (SFH) modeling inferences, we show the results from an example clump across the primary lensed arc in \twentyoneeleven, sampled by both NIRSpec/prism spectroscopy and NIRCam imaging. 

We define the Balmer break as the ratio of the median observed flux density measured redward and blueward of the Balmer break:
\begin{equation}
    BB
    \equiv \frac{\widetilde{F}_{\nu,\,\mathrm{red}}}{\widetilde{F}_{\nu,\,\mathrm{blue}}}
    = \frac{\mathrm{median}\left[ F_\nu(4150\text{\AA} \le \lambda \le 4250\text{\AA}) \right]}
           {\mathrm{median}\left[ F_\nu(3400\text{\AA} \le \lambda \le 3600\text{\AA}) \right]} \, .
\end{equation}
where $\widetilde{F}_{\nu,\,\mathrm{red}}$ and $\widetilde{F}_{\nu,\,\mathrm{blue}}$ denote the median observed flux densities in the wavelength intervals [4150, 4250]\,\AA\ and [3400, 3600]\,\AA, respectively. A higher value, or a stronger Balmer break, is associated with old stellar populations and/or past quenching of star formation (see \citealt{mintz2026} and references therein).

\begin{figure*}[htb!]
    \centering
    ~\hspace{-7mm}\includegraphics[width=1.05\textwidth]{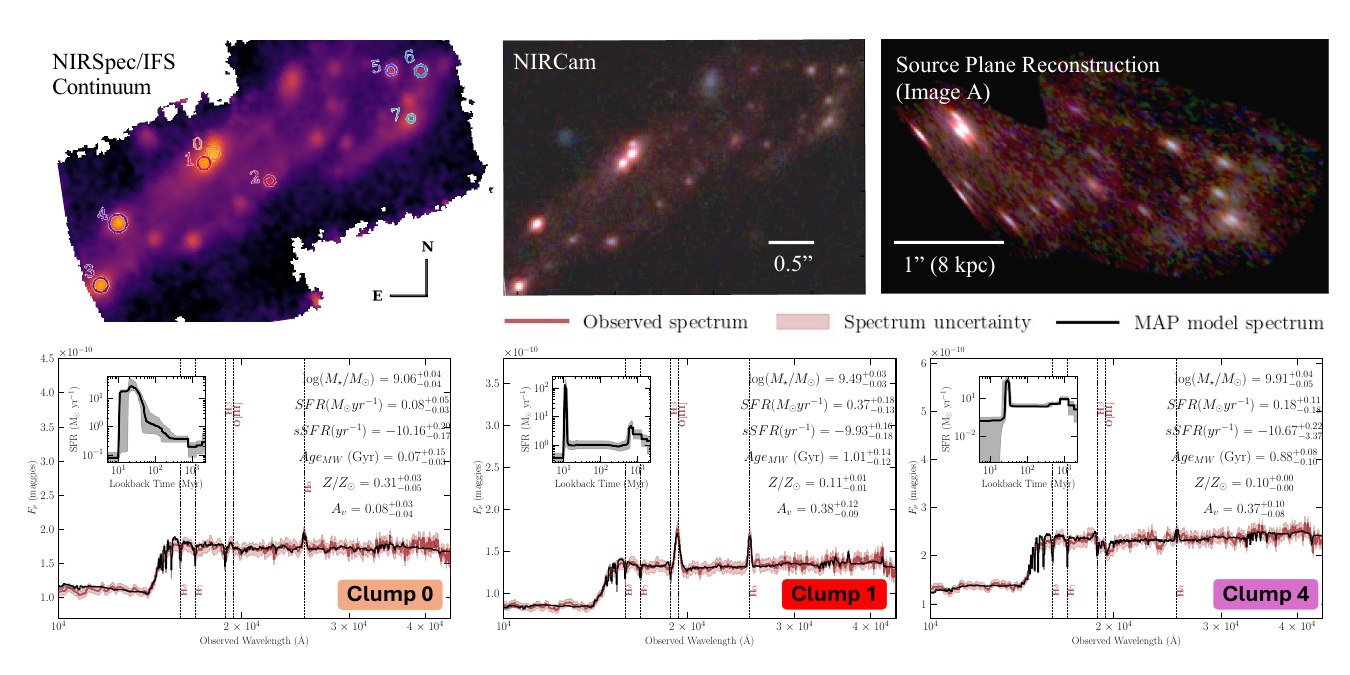}
    \caption{SED Fitting example of quenched ``clumpy" regions within SGAS2111 - Clump 0, 1 and 4 (see Figure \ref{fig:joyplot_spectra_sgas2111}). This spectrophotometric fit (non-smoothed FSPS-based \code{Prospector} model in black) shows clumps that underwent a starburst phase 50--150 Myr before the epoch of observation, followed by a quenching episode. The spectrum of these region within SGAS2111 (spectral data in red, uncertainty in orange) contains no (or weak) emission lines, strong Balmer breaks, and an upturn in UV flux; this is reminiscent of high-redshift post-starburst galaxies seen in recent JWST literature -- nappers. Please note that stellar masses/star formation rates are not lensing-corrected.} 
    \label{fig:sed_clump}
\end{figure*}

Figure \ref{fig:2111_bb_ha} shows the stellar continuum, $H_{\alpha}$ and Balmer break maps for SGAS2111, showing co-spatial and distinct star-forming vs quenched regions within the galaxy. Figure \ref{fig:joyplot_spectra_sgas2111} used seven of these clumpy regions to extract 1D spectra (plotted in decreasing order of Balmer Break strength). Clumps 4, 1 and 0 -- the three brightest regions in the galaxy -- have some of the largest Balmer breaks observed, which we confirm with \code{Prospector}-based SPS modeling (shown in Figure \ref{fig:sed_clump}).  We show the NIRSpec/PRISM spectra (in red) with best-fitting models (in black); the star formation history (SFH, inset plot) shows a recent burst and a subsequent quenching episode (100 and 20 Myr before the epoch of observation, respectively) -- clearly representative of a component of this galaxy not actively forming stars, matching the lack of emission lines in observed spectra as well. 

Clumps 0 and 1 are adjacent to each other (200-400pc), while clump 4 is $\sim$ 2 kpc away, which indicates a quenching mechanism that is relatively global in the north-west region of the galaxy/galaxies (see the source plane reconstruction for SGAS2111 in the right panel of Figure \ref{fig:sed_clump}), potentially a product of multiple merger episodes, given the field of SGAS2111 is $\sim$ 15-25 kpc wide; this is a by-product of observing a galaxy like SGAS2111 that is both radially and tangentially extended in the lensed image plane (and hence is expected to be a field that spans multiple kiloparsecs in the source plane). Both the older quenching epoch and the lower sSFR for clump 4 indicates that it quenched first (and for longer) relative to clumps 0 and 1. Also note that clump 0 is a relatively younger and dust-free region that was actively forming stars from a Gyr to 50 Myr before the epoch of observation, yet quenched at the same time as clump 1. 

The detailed SPS Modeling for this interacting system filled with quenched regions will be conducted in an upcoming publication (G. Khullar et al. in prep, 2026).

\subsection{Sunburst Arc in the Literature}

Within a decade of discovery, the Sunburst Arc has already become a powerhouse laboratory for detailed, high-resolution studies of stellar populations and resolved ISM properties at Cosmic Noon. This section summarizes the body of work on this lensed arc, both outside and within the scope of LEGGOS.

The Sunburst Arc ($z=2.37006$) was discovered by \cite{dahle2016}; its source is lensed by the massive galaxy cluster PSZ1G311.65$-$18.48 ($z=0.443$), which was initially found through the Sunyaev-Zeldovich effect in the Planck data by the Planck Foreground project \citep{plancksz2014}. 
Rest-UV follow-up spectroscopy with Magellan/MagE revealed a unique, triple-peaked Lyman-\(\alpha\) line profile \citep{sunburst2017,owens2024} which simulations have predicted as a tell-tale sign of Lyman-continuum (LyC) emission through a narrow escape path of extremely low \ion{H}{1} column density \citep[e.g.][]{behrens2014,verhamme2015}. This prediction was confirmed when subsequent \textit{Hubble} Space Telescope WFC3/F275W imaging showed bright, but extremely compact, LyC emission \citep{Rivera-Thorsen_2019}, with an absolute LyC escape fraction of \(f_{\text{esc,abs}}^{\text{LyC}} = 0.35 \pm 0.15\).  
Lyman-\(\alpha\) radiative transfer modeling originally suggested that in order to give rise to a triple-peaked Ly\(\alpha\) profile such as observed in the Sunburst Arc, the empty escape path of the middle peak must have a very low opening angle \citep[\(\lesssim 2-5\%\) of the total solid angle;][]{behrens2014,sunburst2017}. However, later and more detailed radiative transfer models have relaxed this requirement significantly, predicting that such a profile could still arise from an opening angle as large as \(\sim 30\%\) of the total solid angle and alleviating the need to invoke uncomfortably fine-tuned scenarios \citep{almadamonter2024}. 

\cite{berg2025} found, using \textit{Hubble} slitless spectroscopy, that the escaping LyC was absorbed below 
\(\lambda_{\text{rest}} \approx 800 \AA\) 
due to an intervening Lyman-limit system, suggesting that the ionizing escape fraction from the Sunburst Arc is likely larger than found by \cite{Rivera-Thorsen_2019}. \cite{berg2025} also demonstrated how the multiply-imaged LyC source could be used to perform neutral IGM tomography along lines at separations lower than previously seen, as suggested by \cite{Rivera-Thorsen_2019}. 

The combination of strong gravitational lensing and a redshift in the interval \(2 \lesssim z \lesssim 4\) has rendered the Sunburst Arc a unique opportunity to study the physical properties that regulate the escape of LyC in very high spatial detail \citep[e.g.][]{Rivera-Thorsen_2019}. 
\cite{mainali2022} found from ground-based rest-optical spectra of the LCE cluster that strong emission lines in [\ion{O}{3}] contained a moderately broad component, similar to low-\(z\) leakers in the literature, and suggested that this might reflect a presence of strong bulk outflows which might have driven the perforation and fracturing of the \ion{H}{1} envelope around the LCE cluster. 
\cite{kim2023} showed from HST imaging that the LCE cluster had a strong UV slope \(\beta_{\text{UV}} = -2.9 \pm 0.1\), consistent with leakers from low-redshift studies \cite[e.g.][]{izotov2018,flury2022}. 
Leveraging the resolution and depth of the JWST/NIRSpec Integral Field Unit, \cite{riverathorsen2025arxiv} found that both dust geometry and strong emission line kinematics both suggested photoionization rather than mechanical feedback as the main generator of the ionizing escape paths. Together with the extremely low neutral column density along the line of sight to the LCE cluster, these authors suggest a scenario where an interaction with a companion galaxy, previously identified by \cite{Sharon2022sunburst}, has stripped away most of the \ion{H}{1} near the LCE in a manner similar to what has been directly observed in the low-redshift LCE galaxy Haro 11 \citep{lereste2024}. Such a tidal stripping would make it possible for the extremely strong emission from the cluster to photoionize the remaining gas to the exceptionally high degree observed. 

The stellar emission from the Sunburst Arc is dominated by the Lyman-Continuum emitting cluster: a young, massive star cluster with a stellar and dynamical mass of just under \(M_{\star} \le 10^7 M_{\odot}\) \citep{vanzella2020a,mestric2023,riverathorsen2024} and a stellar age of 3-4 Myr \citep{chisholm2019, mestric2023, pascale2023, riverathorsen2024}. The cluster shows indirect evidence of a population of very massive stars (VMC; \citealt{mestric2023}), and a sizable population of classical Wolf-Rayet (WR) stars \citep{riverathorsen2024}, one of the highest redshift WR populations detected. 
The LCE cluster has shown signs of a strongly elevated N/O abundance \citep{pascale2023,welch2025}, placing it in a class of so-called Nitrogen-loud galaxies, together with e.g. GN-z11 \citep{ji2024,vink2023}. This supports the hypothesis of \cite{ji2024} that much of such strong N/O build-up is found close to massive, evolved stars in the most massive clusters in these galaxies.

The Sunburst Arc is host to a bright, peculiar feature, referred to in the literature as ``Tr'' \citep{vanzella2020}, the ``peculiar clump'' \citep{Sharon2022sunburst}, or ``Godzilla'' 
\citep{diego2022, pascale2024, choe2025}. Although the source galaxy is fully or partially imaged as many as 12 times along the arc, this clump appears to have no counterimages \citep{pignataro2021,diego2022, Sharon2022sunburst, solhaug2025}. 
Spectroscopically, the clump also shows highly unusual features, including very strong fluorescent emission lines pumped by accidental resonance with \ion{H}{1} Ly\(\;\alpha\) and Ly\(\;\beta\) \citep{vanzella2020,choe2025}. 
\cite{vanzella2020} suggested that the feature is a transient, most likely a supernova, which would account for the lack of counterimages with the strong fluorescent emission lines originating in a region where the SN remnant was interacting with the surrounding ISM. 
However, \cite{Sharon2022sunburst} demonstrated that the SN interpretation was inconsistent with the time delays expected from the lens model. \cite{diego2022} posited that the object is a single Luminous Blue Variable (LBV) star, magnified to an extreme amount due to a fortuitous alignment with the critical curves in the lens system in a manner similar to what has previously been observed in, e.g., the lensed star candidate Earendel \citep{welch2022}; and they nicknamed the feature ``Godzilla''. 
\cite{pascale2024} argue that ``Godzilla'' is instead a star cluster, subject to strong but not extreme magnification, and suggested a number of fainter candidate counterimages. 
\cite{choe2025} challenged both the cluster interpretation of \cite{pascale2024} and the single-star interpretation of \cite{diego2022}. Based on the PID 02555 NIRCam images, they showed that the colors of ``Godzilla'' are inconsistent with the candidate counterimages from \cite{pascale2024}; and from combining this imaging with NIRSpec integral field spectroscopy, they could also rule out the single-star suggestion from \cite{diego2022}. Instead, \cite{choe2025} proposed a hybrid scenario, in which ``Godzilla" is an extremely bright and magnified LBV-type star similar to \(\eta\) Car, embedded in a star cluster.

Building on this extensive body of work, the LEGGOS analyses will unpack the nature of the complete sample of clumps/star clusters across Sunburst, with the combination of NIRCam and NIRSpec coverage -- both within the context of Sunburst, but also in relation to the physics of clumpy star formation across all systems beyond Cosmic Noon.

\section{Summary} \label{sec:summary}

In this work, we introduce The LEGGOS Survey (LEnsing and Galaxy Growth: Observing Substructures), a JWST program using strong gravitational lensing with NIRCam imaging and NIRSpec IFU spectroscopy to resolve clumpy star formation and quenching in eight UV-bright gravitationally lensed galaxies at z = 2--4. The eight survey targets --- SGAS1110, SGAS1527, SGAS1050, SGAS2111, SGAS1429, the Cosmic Eye, the Sunburst Arc, and SGAS1226 --- are highly magnified arcs reaching $\sim$ 10--200 pc source-plane resolution in rest-frame optical continuum and nebular emission, enabling clump-scale measurements that are inaccessible in unlensed field samples to date.

LEGGOS combines Cycle 2 observations (GO 4125/3843) with archival datasets (including the Sunburst Arc and SGAS1226) and delivers a uniform workflow that couples lens reconstruction, PSF-matched multi-band clump photometry, Bayesian spectrophotometric SPS modeling (\texttt{Prospector} with flexible nonparametric SFHs and dust prescriptions), and resolved emission-line diagnostics from emission line spectroscopy. The spectroscopic component provides Balmer-line constraints on attenuation and SFR and uses strong-line ratios to probe ionization and gas-phase abundance, explicitly breaking degeneracies inherent to photometry-only clump studies.

Through early results, we demonstrate the power of LEGGOS. Spatially resolved [N II]--BPT diagnostics in SGAS1110 and the Cosmic Eye reveal a diversity of ionizing conditions within individual galaxies, with SGAS1110 dominated by high-ionization star-forming regions ($\sim$80\%) and the Cosmic Eye exhibiting predominantly composite ratios ($\sim$73\%). N2-based metallicity maps for three targets show qualitatively distinct enrichment structures: SGAS1110 has a broad distribution consistent with clump-to-clump chemical variation; the Cosmic Eye shows a uniform, near-Gaussian metallicity profile indicative of efficient metal mixing; and SGAS1429 exhibits elevated metallicity in multiply-imaged clumps relative to diffuse emission, possibly reflecting older, more evolved stellar populations. Moreover, non-parametric SFH modeling of quenched clumps in SGAS2111 with \code{Prospector} reveal regions with starburst episodes $\sim$50--150 Myr before observation, followed by rapid quenching $\sim$20 Myr prior --- a post-starburst ``napper'' SFH with no nebular emission, a strong Balmer break, and a rest-frame UV upturn. The co-existence of recently ``quenched'' and some actively star-forming clumps within a single galaxy at z=2.858 --- spatially located a few kpc from each other --- provides an example for spatially heterogeneous mass assembly at cosmic noon, likely via mergers.
    
Forthcoming publications from LEGGOS will present full clump SPS modeling across the sample, a stellar mass function characterization, and spatially resolved line diagnostics across all spaxels, among others. We expect LEGGOS --- both the dataset and methodological frameworks --- to be a benchmark in the studies of clumpy star formation and cessation in this new era of utilising the combination of \JWST and gravitational lensing.

\acknowledgments 

GK acknowledges that a significant part of our work is done on stolen land, and we support the efforts of movements like \#LandBack and passing on true stewardship to those same peoples whose land is occupied. The authors also note that we do not use the full name of JWST in the publications we share, due to the person after whom this telescope is named, and their role as NASA administrator during the ``Lavender Scare'', as per the \#RenameJWST protest movement. 

GK would like to acknowledge the support of Jessica Werk, Matthew McQuinn, Jessica Ness and Liza Young in the construction and management of the \jwst LEGGOS survey. GK thanks Arianna Long for stimulating discussions on the molecular gas and cold dust contents in LEGGOS galaxies, and Jed McKinney for sharing the joys of a joyplot (see Figure \ref{fig:joyplot_spectra_sgas2111}). GK is indebted to NIRCam and NIRSpec astronomers and engineers at STScI who helped us develop the observational strategy (including the Telescope Time Review Board -- TTRB -- for iterating with us on efficient usages of telescope resources). GK would like to thank the Baum Grant and Fellowship at the University of Washington for support during this work, as well as the ALMA Ambassador Program (administered by NAASC and NRAO). GK also thanks the International Space Science Institute (ISSI), Bern, for their hospitality, financial support and collaboration during the time of writing this manuscript. GK would also like to thank the DiRAC Institute in the Department of Astronomy at the University of Washington. The DiRAC Institute is supported through generous gifts from the
Charles and Lisa Simonyi Fund for Arts and Sciences, Janet and Lloyd Frink, and the Washington Research Foundation. 

This research was supported in part by the University of Pittsburgh Center for Research Computing, RRID:SCR\_022735, through the resources provided. Specifically, this work used the H2P/MPI cluster, which is supported by NSF award number OAC-2117681. 
ER-T is supported by the Swedish Research Council grant 2022-04805.
TAH's research is supported by an appointment to the NASA Postdoctoral Program at the NASA Goddard Space, administered by Oak Ridge Associated Universities under contract with NASA, as well as the University of Maryland Baltimore County and the Center for Space Sciences and Technology.

This work is based primarily on observations made with the NASA/ESA/CSA \emph{JWST}. The data were obtained from the Mikulski Archive for Space Telescopes at the Space Telescope Science Institute, which is operated by the Association of Universities for Research in Astronomy, Inc., under NASA contract NAS 5-03127 for \JWST. These observations are associated with \JWST Cycle 2 GO programs \#4125, 3843 and archival data from GO 2555 and JWST-ERS-01355. The specific observations analyzed can be accessed via \dataset[https://doi.org/10.17909/70dh-0x38]{https://doi.org/10.17909/70dh-0x38} and \dataset[https://doi.org/10.17909/h90f-n539]{https://doi.org/10.17909/h90f-n539}.

Support for the GO programs was provided by NASA through a grant from the Space Telescope Science Institute, which is operated by the Associations of Universities for Research in Astronomy, Incorporated, under NASA contract NAS5-26555. This research is based on observations made with the NASA/ESA Hubble Space Telescope obtained from the Space Telescope Science Institute, which is operated by the Association of Universities for Research in Astronomy, Inc., under NASA contract NAS 5-26555. New HST observations are associated with programs HST-GO-17311; see appendix for HST program information for archival observations used in this work. Support to MAST for these data is provided by the NASA Office of Space Science via grant NAG5–7584 and by other grants and contracts.

\vspace{5mm}
\facilities{\JWST{}(NIRCam, NIRSpec, and MIRI), \HST{}(ACS and WFC3)}

\software{\code{Python 3.10 --- Prospector \citep{Johnson2021a}, python-FSPS \citep{Conroy2009,Conroy2010}, SEDpy, Matplotlib \citep{Hunter2007}, Numpy \citep{numpy2020}, Scipy \citep{scipy2020}, Astropy \citep{astropy2013, astropy2018}, Jupyter, IPython Notebooks, DrizzlePac \citep{Gonzaga2012}}, CubeFitter.jl \citep{cubefitter2025}.}


\bibliography{all}

\appendix

\section{Foundations of a JWST Survey: Burnout and Joy}

The JWST LEGGOS Survey began in 2023 as a collaboration led by a group of predominantly junior (non-tenured) astronomers supported by senior faculty and research staff across the world. In order to build a supportive and inclusive team  to work through data collection, analysis, and circulation of data products and publications, we explored the possibility of embedding foundational concepts around ``burnout" and ``joy" within the functioning of the team. 

In practice, the LEGGOS collaboration designed all decision-making structures, policies, and science projects to simultaneously take into account the scientific interests of lead authors/co-authors as well as the team's mental, physical, and emotional bandwidth around the breadth of science explored by the collaboration. In addition to science discussions, at each team meeting (whether in-person or hybrid) we feature discussions about community, inclusion, and justice, centering marginalized and junior astronomers in LEGGOS (e.g., discussions around burnout and joy, inspired from the work of sociologist, organizer, and author Dean Spade and his book ``Mutual Aid: Building Solidarity During this Crisis (and the Next)'' \citep{spade2020mutual}, which presented a framework for identifying burnout and focusing on joy within teams). 

The details on the motivation and praxis of this approach will be laid out in a separate publication (Khullar, Hutchison et al. in prep, 2026). 

\section{Ethos behind The Survey Logo}

The JWST LEGGOS Survey has a logo that embodies the spirit of the observational programs and its science goals, while placing an emphasis on the targets from GO-4125 and GO-3843, and the ethos behind the team/collaboration's work ethic as part of the team.\footnote{Which includes the intentional choice to submit this work during LGBTQIA+ Pride Month (in the United States).} See Figure \ref{fig:mosaic} for one of the versions of the logo, as well as the open-source (and shareable) version here: \href{https://zenodo.org/records/20721247}{Zenodo Link} \footnote{Available at \url{https://zenodo.org/records/20721247} under the CC-BY-NY-SA open-source license.}. 

\section{Pre-JWST Observations of LEGGOS Lensed Galaxies}

Here, we consolidate archival observations -- both ground- and space-based -- for galaxies in our sample, for the reader to utilize, in the context of JWST as well as distinct from it. Each table in the appendix corresponds to a single galaxy in the LEGGOS sample.

\begin{deluxetable*}{lp{1.4in}p{1.0in}p{0.9in}p{2.0in}}
\tabletypesize{\scriptsize}
\tablecaption{Sunburst Observations}
\label{tab:sunburst}
\tablehead{
  \colhead{Facility/instrument} & \colhead{Type of observation} & \colhead{PID/PI} & \colhead{Reference(s)} & \colhead{Comments}
}
\startdata
HST/ACS+WFC3 & Wide+medium band imaging + IR grism & 15101 Dahle & \citet{Rivera-Thorsen_2019} & F410M+F555W+F814W+F105W+F140W, G141 \\
HST/WFC3 & Wide+medium band imaging & 15377 Bayliss & \citet{Sharon2022sunburst} & F606W+F098M+F125W+F160W \\
HST/WFC3 & LyC imaging & 15418 Dahle & \citet{Rivera-Thorsen_2019} & 3 orbits F275W; discovery of LyC leakage \\
HST/WFC3 & LyC + narrow/medium-band em.line imaging & 15949 Gladders & \citet{kim2023} & F275W (26 orbits), F390W, F126N, F128N, F140W, F153M, F164, F167N \\
HST/WFC3 & LyC UV grism spectroscopy & 15966 Chisholm & \citet{berg2025} & ... \\
NTT/EFOSC2 & R+z band imaging & 192.A-0762 Aghanim & \citet{dahle2016} & (Discovery data) \\
Magellan/IMACS & Optical spectroscopy & Bayliss & \citet{dahle2016} & (Redshift confirmation) \\
Magellan/4star & IR imaging & Bayliss & \citet{dahle2016} & J+Ks imaging \\
Magellan/MagE & Optical spectroscopy & Bayliss & \citet{sunburst2017,chisholm2019,mainali2022,owens2024} & Multiple pointings, both LyC images and non-leaker regions \\
Magellan/MagE & Optical spectroscopy & CN2017B-57  Tejos & \citet{lopez2020} & Three pointings with slit along arc \\
Magellan/FIRE & IR spectroscopy & Bayliss & \citet{sunburst2017,mainali2022} & Multiple pointings, both LyC images and non-leaker regions \\
Magellan/MIKE & hi-res optical spectroscopy & Bayliss & \citet{solhaug2025} & Multiple pointings, both LyC images and non-leaker regions \\
VLT/FORS2 & MOS spectroscopy of lens cluster members & 192.A-0762 Aghanim & ... & ... \\
VLT/MUSE (1) & optical IFU spectroscopy & 297.A-5012(A) Aghanim & \citet{lopez2020,vanzella2020,pignataro2021,Sharon2022sunburst} & 1.2h integration DDT programme \\
VLT/MUSE (2) & optical IFU spectroscopy & 0103.A-0688(A) Vanzella & \citet{vanzella2020} & LyC leaker image 11 \\
VLT/MUSE (3) & optical IFU spectroscopy & 2107.D-5057(A), 0110.B-0242(A) Vanzella & \citet{mestric2023} & deep (7h in total) \\
VLT/KMOS & IR MOS spectroscopy & 0105.A-0387(A) Vanzella & ... & ... \\
VLT/ESPRESSO & hi-res optical spectroscopy & 60.A-9507(A) Vanzella & \citet{vanzella2020} & (Detection of Bowen fluorescence in "Godzilla") \\
VLT/XSHOOTER & optical + NIR spectroscopy & 0103.A-0688  Vanzella & \citet{vanzella2020} & (Detection of Bowen fluorescence in "Godzilla") \\
VLT/UVES & hi-res optical spectroscopy & 0113.B-2084(A) Lopez & ... & ... \\
Spitzer & IR imaging & P13111 Dahle & ... & 3.6, 4.5 micron imaging \\
ALMA/ACA & mm-wave & 2018.1.01142.S González-López & \citet{solimano2021} & non-detection \\
AstroSat UVIT & far-UV imaging & A07\_165 Rivera-Thorsen & ... & (non detection in F154W) \\
Chandra ACIS-I & X-ray & 19800436 Bayliss & \citet{Gassis26a} & Obs ID 20567 and 22081; see also \citet{Gassis26b} \\
\enddata
\tablecomments{Observations are listed chronologically by instrument. 
PID/PI: Program ID and Principal Investigator. 
References provided in parentheses; see bibliography for full citations.}
\end{deluxetable*}

\begin{deluxetable*}{lp{1.4in}p{1.0in}p{0.9in}p{2.0in}}
\tabletypesize{\scriptsize}
\tablecaption{SGASJ1050 Observations}
\label{tab:sgasj1050}
\tablehead{
  \colhead{Facility/instrument} & \colhead{Type of observation} & \colhead{PID/PI} & \colhead{Reference(s)} & \colhead{Comments}
}
\startdata
HST WFC3 & UVIS+IR wide-band imaging & 13003 Gladders & \citet{Bayliss2014} & F390W, F606W, F110W, F160W; see also Sharon+20 \\
HST/ACS & Ramp filter narrow-band imaging & 13639 Bayliss & \citet{Navarre24} & FR716N; captures LyA for secondary arc @ z=4.867 \\
Magellan/IMACS & mask spectroscopy & Bayliss & \citet{Bayliss2014} & See also \citet{Bayliss14} Two masks, confiming additional image of main arc and targeting other candidate lensed sources in the field \\
Magellan/MagE & Optical spectroscopy & Rigby & \citet{Bayliss2014} & Single slit; see \citet{Bayliss2014} Fig. 2 for slit position, also in MegaSaura \citep{Rigby_2018a} \\
Magellan/FIRE & IR spectroscopy & Rigby & \citet{Bayliss2014} & Single slit; see \citet{Bayliss2014} Fig. 2 for slit position \\
Subaru/Suprime-Cam & broadband imaging & Oguri & \citet{Oguri12} & joint strong + weak lensing analysis from grz imaging \\
Gemini/GMOS & N\&S mask spectroscopy & Sharon & \citet{Bayliss2014} & See also \citet{Bayliss14} Three mask slits along arc; see \citet{Bayliss2014} Fig. 2 for slit locations \\
VLT/FORS2 & imaging & 087.A-0539(B) Barrientos & ... & I-band imaging \\
VLT/HAWK-I & IR imaging & 087.A-0539(A) Barrientos & ... & H-band imaging \\
VLT/SINFONI & IFU spectroscopy & 092.A-0095(A) Wuyts & ... & ... \\
Spitzer/IRAC & IR imaging & P70154 Gladders, P90232 Rigby & \citet{Bayliss14} & ... \\
ACA & mm-wavelength & 2022.1.00916.S + 2023.1.01011.S & \citet{catan2024} & dust (detected) and CO (upper limit) \\
Chandra ACIS-I & X-ray & 19800436 Bayliss & \citet{Gassis26a} & Obs ID 20567 and 22081; see also \citet{Gassis26b} \\
NOT/MOSCA & imaging & Dahle & ... & SGAS1 grz imaging (March 14, 2010); initial confirmation of lens candidate from SDSS \\
\enddata
\tablecomments{Observations are listed chronologically by instrument. 
PID/PI: Program ID and Principal Investigator. 
References provided in parentheses; see bibliography for full citations.}
\end{deluxetable*}

\begin{deluxetable*}{lp{1.4in}p{1.0in}p{0.9in}p{2.0in}}
\tabletypesize{\scriptsize}
\tablecaption{SGASJ1110 Observations}
\label{tab:sgasj1110}
\tablehead{
  \colhead{Facility/instrument} & \colhead{Type of observation} & \colhead{PID/PI} & \colhead{Reference(s)} & \colhead{Comments}
}
\startdata
HST WFC3 & UVIS+IR wide-band imaging & 13003 Gladders & \citet{Johnson2017_1110paperI} & see also \citet{johnson2017apjl_1110paperIII}, \citet{rigby2017_1110paperII}  F390W, F606W, F105W, F160W; see also \citet{Sharon20} \\
MMT/BCS & spectroscopy & ... & \citet{Johnson2017_1110paperI} & Longslit along the arc \\
MMT/BCS & spectroscopy & ... & \citet{Stark_2013} & CASSOWARY, longslit crosses arc + BCG \\
Subaru/Suprime-Cam & broadband imaging & Oguri & \citet{Oguri12} & joint strong + weak lensing analysis from grz imaging \\
Gemini/GMOS & N\&S mask spectroscopy & GN-2011A-Q-19 (PI: Gladders) and GN-2015B-Q-26 (PI: Sharon) & \citet{Johnson2017_1110paperI} & Cluster member + arc spectroscopy \\
Spitzer/IRAC & IR imaging & P70154 Gladders, P90232 Rigby & \citet{Johnson2017_1110paperI} & ... \\
Chandra ACIS-I & X-ray & 19800436 Bayliss & \citet{Gassis26a} & Obs ID 20568 and 22099; see also \citet{Gassis26b} \\
NOT/MOSCA & imaging & Dahle & ... & SGAS1 grz imaging (March 14, 2010); initial confirmation of lens candidate from SDSS \\
\enddata
\tablecomments{Observations are listed chronologically by instrument. 
PID/PI: Program ID and Principal Investigator. 
References provided in parentheses; see bibliography for full citations.}
\end{deluxetable*}

\begin{deluxetable*}{lp{1.4in}p{1.0in}p{0.9in}p{2.0in}}
\tabletypesize{\scriptsize}
\tablecaption{SGASJ1226 Observations}
\label{tab:sgasj1226}
\tablehead{
  \colhead{Facility/instrument} & \colhead{Type of observation} & \colhead{PID/PI} & \colhead{Reference(s)} & \colhead{Comments}
}
\startdata
HST/WFPC2 & wide-band imaging & 11103 Ebeling & \citet{Tejos21} & F606W SNAP, shallow (1200s total) \\
HST/WFC3 & UVIS wide-band imaging & 12368 Morris & \citet{Tejos21} & F606W, F814W; see also \citet{sharon2022} \\
HST/ACS & wide-band imaging & 12368 Morris, 12166 Ebeling & \citet{Tejos21} & F606W, F814W; see also \citet{sharon2022} \\
HST/WFC3 & IR wide-band imaging & 12166 Ebeling, 15378 Bayliss & \citet{Tejos21} & F110W, F140W, F160W; see also \citet{sharon2022} \\
Gemini/GMOS & N\&S mask spectroscopy & ... & \citet{Bayliss11} & ... \\
Gemini/GMOS & imaging & ... & \citet{Koester_2010} & ... \\
APO3.5 & IR imaging & ... & \citet{Wuyts2012} & JHKs \\
MMT/Hectospec & MOS spectroscopy & ... & \citet{Bayliss14} & ... \\
Magellan/MagE & slit spectroscopy & ... & \citet{Koester_2010} & Deeper MagE spectrum in MegaSaura, see \citet{Rigby_2018a} \\
Magellan/IMACS+GISMO & MOS spectroscopy & ... & \citet{Bayliss14} & ... \\
Keck/NIRSPEC & IR spectroscopy & ... & \citet{Wuyts2012} & ... \\
Subaru/Suprime-Cam & broadband imaging & Oguri, Ebeling & \citet{Oguri12} & joint strong + weak lensing analysis from gRz imaging \\
VLT/MUSE & IFU spectroscopy & 0101.A-0364(A) Lopez, 0102.A-0718(B) Edge & \citet{Tejos21} & ... \\
VLT/HAWK-I & IR imaging & 0100.A-0785(A) Edge & ... & Ks-band imaging \\
VLT/XSHOOTER & OIR single-slit spectroscopy & 094.B-0041(A) Swinbank & ... & ... \\
Spitzer/IRAC & IR imaging & P70154 Gladders & \citet{Wuyts2012} & ... \\
ACA & mm-wavelength & 2022.1.00916.S + 2023.1.01011.S & \citet{catan2024} & ... \\
Chandra ACIS-I & X-ray & 19800436 Bayliss & \citet{Gassis26a} & Obs ID 12878; see also \citet{Gassis26b} \\
NOT/MOSCA & imaging & Dahle & ... & SGAS1 g-band imaging (December 11, 2007); initial confirmation of lens candidate from SDSS; subsequent deep gri imaging\\
\enddata
\tablecomments{Observations are listed chronologically by instrument. 
PID/PI: Program ID and Principal Investigator. 
References provided in parentheses; see bibliography for full citations.}
\end{deluxetable*}

\begin{deluxetable*}{lp{1.4in}p{1.0in}p{0.9in}p{2.0in}}
\tabletypesize{\scriptsize}
\tablecaption{SGASJ1429 Observations}
\label{tab:sgasj1429}
\tablehead{
  \colhead{Facility/instrument} & \colhead{Type of observation} & \colhead{PID/PI} & \colhead{Reference(s)} & \colhead{Comments}
}
\startdata
HST WFC3 & UVIS+IR wide-band imaging & 15378 Bayliss & ... & F475X, F814W, F110W, F160W \\
Magellan/MagE & Optical spectroscopy & ... & \citet{Rigby_2018a} & megasaura \\
GTC/OSIRIS & imaging & ... & \citet{Marques-Chaves17} & g-band \\
GTC/OSIRIS & optical spectroscopy & ... & \citet{Marques-Chaves17} & Independent discovery from BELLS follow-up \\
ACA & mm-wavelength & 2022.1.00916.S + 2023.1.01011.S & \citet{catan2024} & ... \\
Chandra/ACIS-I & X-ray & 19800436 Bayliss & \citet{Gassis26b} & Obs ID 20574, 22188 \\
NOT/MOSCA & imaging & Dahle & \citet{Rigby_2018a} & SGAS2 grz imaging (April 18, 2012); initial confirmation of lens candidate from SDSS \\
\enddata
\tablecomments{Observations are listed chronologically by instrument. 
PID/PI: Program ID and Principal Investigator. 
References provided in parentheses; see bibliography for full citations.}
\end{deluxetable*}

\begin{deluxetable*}{lp{1.4in}p{1.0in}p{0.9in}p{2.0in}}
\tabletypesize{\scriptsize}
\tablecaption{SGASJ1527 Observations}
\label{tab:sgasj1527}
\tablehead{
  \colhead{Facility/instrument} & \colhead{Type of observation} & \colhead{PID/PI} & \colhead{Reference(s)} & \colhead{Comments}
}
\startdata
HST/WFC3 & UVIS+IR wide-band imaging & 13003 Gladders & \citet{Sharon20} & F475W, F606W, F110W, F160W \\
Magellan/MagE & Optical spectroscopy & ... & \citet{Koester_2010} & Also note multiple MagE slit positions in MegaSaura; see \citet{Rigby_2018a} \\
Keck/NIRSPEC & IR spectroscopy & ... & \citet{Wuyts2012} & ... \\
APO3.5 & IR imaging & ... & \citet{Wuyts2012} & JHKs \\
Gemini/GMOS & imaging & ... & \citet{Koester_2010} & gri imaging \\
Gemini/GMOS & N\&S mask spectroscopy & ... & \citet{Bayliss11} & ... \\
VLT/MUSE & IFU spectroscopy & 0103.A-0485(B) Lopez & \citet{Lopez_2024} & ... \\
Spitzer/IRAC & IR imaging & P50823 Gladders & \citet{Wuyts2012} & ... \\
ACA & mm-wavelength & 2022.1.00916.S + 2023.1.01011.S & \citet{catan2024} & ... \\
WIYN/OPTIC & imaging & ... & \citet{Hennawi_2008} & May 2005 SGAS1 imaging initial confirmation of lens candidate from SDSS \\
\enddata
\tablecomments{Observations are listed chronologically by instrument. 
PID/PI: Program ID and Principal Investigator. 
References provided in parentheses; see bibliography for full citations.}
\end{deluxetable*}

\begin{deluxetable*}{lp{1.4in}p{1.0in}p{0.9in}p{2.0in}}
\tabletypesize{\scriptsize}
\tablecaption{SGASJ2111 Observations}
\label{tab:sgasj2111}
\tablehead{
  \colhead{Facility/instrument} & \colhead{Type of observation} & \colhead{PID/PI} & \colhead{Reference(s)} & \colhead{Comments}
}
\startdata
HST WFC3 & UVIS+IR wide-band imaging & 13003 Gladders & \citet{Sharon20} & F390W, F814W, F110W, F160W \\
VLT/MUSE & IFU spectroscopy & 0103.A-0485(B) Lopez & \citet{Lopez_2024} & ... \\
Magellan/MagE & Optical spectroscopy & ... & \citet{Rigby_2018a} & MegaSaura \\
Magellan/IMACS & MOS spectroscopy & ... & \citet{Sharon20} & ... \\
Subaru/Suprime-Cam & broadband imaging & Oguri & \citet{Oguri_2009} & See also \citet{Oguri12}; joint strong + weak lensing analysis from grz imaging \\
Gemini/GMOS & N\&S mask spectroscopy & ... & \citet{Bayliss11} & ... \\
Spitzer/IRAC & IR imaging & P70154 Gladders, P90232 Rigby & ... & ... \\
ACA & mm-wavelength & 2022.1.00916.S + 2023.1.01011.S & \citet{catan2024} & ... \\
WIYN/Minimo & imaging & ... & \citet{Hennawi_2008} & SGAS1 gri imaging initial confirmation of lens candidate from SDSS \\
\enddata
\tablecomments{Observations are listed chronologically by instrument. 
PID/PI: Program ID and Principal Investigator. 
References provided in parentheses; see bibliography for full citations.}
\end{deluxetable*}

\begin{deluxetable*}{lp{1.4in}p{1.0in}p{0.9in}p{2.0in}}
\tabletypesize{\scriptsize}
\tablecaption{Cosmic Eye Observations}
\label{tab:cosmic_eye}
\tablehead{
  \colhead{Facility/instrument} & \colhead{Type of observation} & \colhead{PID/PI} & \colhead{Reference(s)} & \colhead{Comments}
}
\startdata
HST/ACS & wide-band imaging & 10491 Ebeling & \citet{Smail07discoverypaper} & F606W (HST SNAP program for MACS clusters); mass model in \citet{dye2007cosmic} \\
HST/ACS & wide-band imaging & 12884 Ebeling & \citet{zitrin2016} & F814W \\
HST/WFC3 & IR imaging & 12166 Ebeling & \citet{zitrin2016} & F110W, F140W \\
Keck/LRIS & long-slit spectroscopy & ... & \citet{Smail07discoverypaper} & Two slit positions covering both arcs \\
Keck/NIRSPEC & IR spectroscopy & ... & \citet{Smail07discoverypaper} & Single slit posn. \\
Keck/NIRC & IR imaging & ... & \citet{Smail07discoverypaper} & K'-band imaging \\
Keck/OSIRIS & (LGS AO-assisted) IFU spectroscopy & ... & \citet{Stark08} & Resolved spectroscopy of well-ordered rotating system \\
Magellan/MagE & Optical spectroscopy & ... & \citet{Rigby_2018a} & MegaSaura \\
VLT/SINFONI & IFU spectroscopy & 083.B-0108 & \citet{Richard11} & ... \\
Spitzer/IRAC & IR imaging & DDT program & \citet{Coppin07} & 3.6, 4.5, 5.8, 8.0, 24 micron \\
Spitzer/MIPS & IR imaging & Program ID 40817 & \citet{Siana09} & (longer wavelengths) \\
Spitzer/IRS & IR imaging (PUI) and spectroscopy & Program ID 40817 & \citet{Siana09} & PAH emission detection at 3.3, 6.2, 7.7 micron \\
\enddata
\tablecomments{Observations are listed chronologically by instrument. 
PID/PI: Program ID and Principal Investigator. 
References provided in parentheses; see bibliography for full citations.}
\end{deluxetable*}

\end{document}